\documentclass[lettersize,journal]{IEEEtran}
\usepackage{amsmath,amsfonts}
\usepackage{array}
\usepackage[caption=false,font=normalsize,labelfont=sf,textfont=sf]{subfig}
\usepackage{textcomp}
\usepackage{stfloats}
\usepackage{url}
\usepackage{verbatim}
\usepackage{mathtools}
\usepackage{algorithmicx,algorithm,float}

\usepackage{algpseudocode}
\newtheorem{lemma}{Lemma}
\newtheorem{mydef}{Definition}

\usepackage[noadjust]{cite}
\newtheorem{theorem}{Theorem}
\newtheorem{remark}{Remark}

\def\x{\mathbf{x}}
\def\V{V_{(k)}}
\def\R{\mathcal{R}}
\def\E{\mathbb{E}}
\DeclarePairedDelimiter{\ceil}{\lceil}{\rceil}

\def\BibTeX{{\rm B\kern-.05em{\sc i\kern-.025em b}\kern-.08em
    T\kern-.1667em\lower.7ex\hbox{E}\kern-.125emX}}
\usepackage{balance}
\begin{document}
\title{Energy-Efficient Computation Offloading in Mobile Edge Computing Systems with Uncertainties}
\author{Tianxi~Ji,
        Changqing~Luo, Lixing Yu, Qianlong Wang, Siheng Chen, Arun Thapa, and Pan Li
\thanks{T. Ji  and P. Li are with the Department
of Electrical, Computer, and Systems Engineering, Case Western Reserve University, Cleveland,
OH, 44106 USA (e-mail: txj116@case.edu;  lipan@case.edu).}
\thanks{C. Luo is with the Department of Computer Science, Virginia Commonwealth University, Richmond, VA 23284 USA (e-mail: cluo@vcu.edu).}
\thanks{L. Yu is with School of Information Science and Engineering, Yunnan University, Yunan, 650500 China (e-mail: yulixing@ynu.edu.cn).}
\thanks{Q. Wang is with Department of Computer and Information Sciences, Towson University, Towson, MD 21252 (e-mail:qwang@towson.edu).}
\thanks{S. Chen is with the School of Electronic Information and Electrical Engineering, Shanghai Jiao Tong University, Shanghai, 200240 China (e-mail: sihengc@sjtu.edu.cn).}
\thanks{A. Thapa is with the Department of Electrical Engineering, Tuskegee University, Tuskegee, AL 36088 USA (e-mail: athapa@tuskegee.edu).}}


\maketitle

\begin{abstract}
\footnote{Please note that a peer-reviewed version of this
paper will be published at IEEE Transactions on Wireless Communications. Copyright will be transferred without notice. There are some minor
editorial differences between the two versions.}Computation offloading is indispensable for mobile edge computing (MEC). It uses edge resources to enable intensive computations and save energy for resource-constrained devices. Existing works generally impose strong assumptions on radio channels and network queue sizes. However, practical MEC systems are subject to various uncertainties rendering these assumptions impractical. In this paper, we investigate the energy-efficient computation offloading problem by relaxing those common assumptions and considering intrinsic uncertainties in the network. Specifically, we minimize the worst-case expected energy consumption of a local device when executing a time-critical application modeled as a directed acyclic graph. We employ the extreme value theory to bound the occurrence probability of uncertain events. To solve the formulated problem, we develop an $\epsilon$-bounded approximation algorithm based on column generation. The proposed algorithm can efficiently identify a feasible solution that is less than $(1+\epsilon)$ of the optimal one. We implement the proposed scheme on an Android smartphone and conduct extensive experiments using a real-world application. Experiment results corroborate that it will lead to lower energy consumption for the client device by considering the intrinsic uncertainties during computation offloading. The proposed computation offloading scheme also significantly outperforms other schemes in terms of energy saving.
\end{abstract}


\section{Introduction}
\label{intro}


%
%
%
%

\IEEEPARstart{T}{he} breakthroughs of hardware and software technologies have resulted in the surge of various mobile applications, such as real-time online gaming and anywhere anytime online social interactions. Mobile devices (e.g., smartphones and wearable devices) executing these applications are usually computation-constrained and battery-limited \cite{satyanarayanan1996fundamental}, which can hinder the development of mobile applications. To address this issue, academia and industry propose to employ  moible edge computing (MEC)   to let mobile devices   offload their computations to edge servers with abundant resources in the proximity \cite{mao2017mobile,wang2020thirty}. As a result, mobile devices are able to support computation-intensive and energy-hungry mobile applications in MEC systems.

So far, researchers have conducted extensive works on energy-efficient computation offloading schemes \cite{geng2018energy,wang2016mobile,zhang2016energy,yu2016joint,salinas2016efficient,sardellitti2015joint,kao2017hermes,luo2017energy,jia2014heuristic,zhang2015collaborative,mahmoodi2016optimal,liu2016delay,wu2019efficient}. Specifically, some works are 
based on   physical layer design of wireless communications \cite{wang2016mobile,sardellitti2015joint}. Some   designed schemes from the perspective of cross-layer design \cite{barbarossa2013computation,zhang2016energy,yu2016joint}. Some   focused on designing offloading schemes for time-critical   applications \cite{kao2017hermes,jia2014heuristic,zhang2015collaborative,ji2021economy,mahmoodi2016optimal,liu2016delay}. 
Previous studies generally impose strong assumptions on communication channels and network queue sizes; they develop computation offloading schemes based on specific radio channels and network queue models. For example, Zhang et al. \cite{zhang2015collaborative} proposed the one-climb policy to offload computations   under independent and identically distributed  (i.i.d.) stochastic channels. Geng et al. \cite{geng2018energy} proposed a critical path based solution to allocate computation tasks to a mobile multi-core device and an edge server without considering dynamic network queues. 

However, a practical MEC system is always subject to   intrinsic uncertainties like unknown communication channels and network queue size.   Factors like weather, obstacles, movements can easily affect radio channels \cite{tse2005fundamentals}. 
For example, it has been shown that the varying locations of the local mobile transmitter and receiver can lead to uncertainties in  channel law as well as arbitrarily varying communication bit rates and  energy consumption \cite{lapidoth1998reliable}. The network queue size is also highly dependent on data volume, network condition, arrival/processing methods, etc., \cite{rappaport1996wireless}. Recent studies have shown that sudden burst of serving requests can overflow the edge servers' resource and cause MEC failure \cite{satria2017recovery,babou2018home}.   
Consequently, it is impossible to use approximated mathematical models to accurately capture the underlying dynamics of radio channel and network queue size \cite{ezio2016impact}. Since  the strong assumptions on radio channels and network queue size cannot hold in practical MEC systems, simply applying computation offloading schemes developed based on the two assumptions can cause   improper parameter deployment at the wireless link layer and even upper layers, which incurs   performance degradation. This is because the parameter deployment at the wireless link layer and upper layers, e.g., transmission and reception power, channel allocation, and routing, are highly related to the radio channels and network queue size. 
Particularly, in  Section \ref{exp:performance_degrade}, we will empirically corroborate the fact that  strong assumptions on  channel conditions and queuing delay will result in much higher  energy consumption at the client device in computation offloading.

In light of this,  we are motivated to investigate the computation offloading problem by taking into account intrinsic uncertainties in practical MEC systems. We model the mobile application as a directed acyclic graph (DAG), and handle the   execution dependency by explicitly considering the parent and children sets of each computation module.  To efficiently solve the formulated the energy-efficient computation offloading problem,  we develop a column generation based $\epsilon$-bounded algorithm with theoretical optimality guarantee. The main contributions of this paper are summarized as follows:

\begin{itemize}
 
 \item We study  energy-efficient computation offloading   by lifting   strong assumptions on communication channels and network queue size imposed by previous studies.
 
 \item We design an energy-efficient computation offloading scheme for executing a mobile application modeled as a DAG in a practical MEC system with uncertainties. Particularly, we employ the extreme value theory to bound the occurrence probability of uncertain events.
 
 \item We formulate the energy-efficient computation offloading problem that is subject to the application execution time and develop an efficient column generation based algorithm to solve it. In particular, we also  provide an $\epsilon$-bounded approximate solution with theoretical guarantee on the optimality of the proposed scheme.
 
 \item We propose computation offloading principles for   mobile applications with only sequential or parallel module dependency in practical MEC systems with uncertainties.
 
 \item We implement our proposed offloading scheme   on an Android platform, and conduct extensive experiments by executing a real-world application, i.e., Smart Diagnosis.  Experiment results corroborate   that we can reduce the energy consumption of the client device by explicitly considering the intrinsic uncertainties in computation offloading, and also demonstrate that our proposed scheme   significantly outperforms other state-of-the-art computation offloading schemes in terms of energy saving. 
 \end{itemize}

\textbf{Roadmap.} We introduce the   related works in Section \ref{RW}. Section \ref{model_mec} describes the considered system model. We present the process of handling the randomness caused by uncertainties of MEC systems in Section \ref{handlernd}, followed by the formulation of the energy-efficient computation offloading problem in Section \ref{PF}. We develop the column generation based algorithm to solve the formulated problem,  and provide computation offloading principles for special mobile applications 
in Section \ref{CG}. We show our experiment results in Section \ref{ex}, and   conclude the paper in Section \ref{con}.

\section{Related Works}
\label{RW}

Quite a few   works have studied the problem of energy-efficient computation offloading in MEC systems \cite{geng2018energy,wang2016mobile,zhang2016energy,yu2016joint,sardellitti2015joint,kao2017hermes,jia2014heuristic,zhang2015collaborative,mahmoodi2016optimal,liu2016delay,hou2020reliable,wang2020thirty}. From the communication perspective, these existing studies can be generally categorized into the physical layer design based, the cross-layer design based, and the time-critical application based. We discuss them as follows.

\subsection{Physical layer design based  schemes.} This line of works   focus on the impact of  electronic circuit transmission on offloading decisions and energy consumption.  For example, Geng et al. \cite{geng2018energy} propose a critical path based solution to recursively check the task modules and move them to the right CPU cores of a multicore  mobile device to save its energy consumption.  
Wang et al. \cite{wang2016mobile} take advantage of dynamic voltage scaling technology to adjust the transmission power at the physical layer in the course of computation offloading. Sardellitti et al. \cite{sardellitti2015joint} jointly optimize the radio resources, constellation size, and CPU cycles per second of the mobile device to minimize the energy consumption of local device.

\subsection{Cross-layer design based   schemes.} This category of works attend to devise offloading systems by exploiting the   resources in different layers in MEC systems. For example, Barbarossa et al. \cite{barbarossa2013computation} propose a joint framework encompassing a fading channel depending on the number of antennas and packet retransmission strategies to determine the joint allocation of radio resources and computation tasks. Zhang et al. \cite{zhang2016energy} incorporate the multi-access characteristics of the 5G heterogeneous network to jointly optimize offloading and radio resource allocation. 
Yu et al. \cite{yu2016joint} minimize the energy consumption of local execution by jointly investigating the subcarrier allocation for offloaded tasks and local CPU time allocation.

\subsection{Time-critical mobile application based   schemes.} These works concentrate on the development of proper delay-sensitive scheduling mechanism  to meet the  quality-of-service  requirements of mobile users
in computation offloading. For example, Hermes \cite{kao2017hermes}, a fully polynomial time approximation scheme, minimizes the application execution latency while meeting the prescribed resource utilization constraints. 
Jia et al. \cite{jia2014heuristic} propose a heuristic programming partition scheme to maximize the parallelism so as to minimize the completion time of an application.  Zhang et al. \cite{zhang2015collaborative} propose a one-climb-policy to offload application with only sequential dependency while meeting a time deadline. 
Mahmoodi et al. \cite{mahmoodi2016optimal}  
apply integer programming to develop a wireless aware joint scheduling and computation offloading to maximize energy saving for devices while satisfying the execution deadline. Liu et al. \cite{liu2016delay} propose a latency-optimal offloading scheme by considering the combination of the queueing state in the task buffer, the execution state of the device's processing unit and transmission unit.  Meng et al. \cite{meng2019closed} use  an infinite horizon average cost Markov
decision process to characterize the interactions between local device and edge server, and develop a   delay-optimal multilevel water-filling computation offloading solution. Hou et al. \cite{hou2020reliable} consider computation offloading for   latency-sensitive applications in Internet of Vehicles.

All these previous studies impose strong assumptions on communication channels and network queue size. However, in practical MEC systems, radio channels and network queue size are usually dynamic and the patterns of the dynamics are unable to be captured accurately \cite{ezio2016impact}.  Particularly, channel status can also be very unpredictable even during the offloading of task modules of an application. As a result, those assumptions cannot hold in real world applications. Although some works, e.g., \cite{kao2017hermes}, adapt an online optimization scheme to continuously probe the radio channels of the unknown dynamic environments, it cannot consider potential extreme cases sufficiently and will introduce extra communication overhead. Therefore, to develop an energy-efficient computation offloading scheme for a practical MEC system is still an open issue. 

\section{System Model}
\label{model_mec}

\subsection{Mobile Application Model}
\begin{figure}[h]
    \centering
    \includegraphics[width=.4\textwidth]{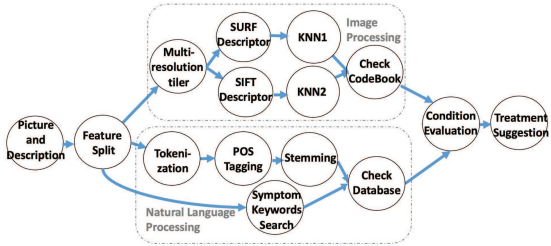}
    \caption{An example of a DAG-modeled mobile application, Smart Diagnosis.} 
    \label{fig:tag}
\end{figure}

We consider a mobile application that is composed of mutually dependent computation task modules (i.e., procedures/components), which need to be executed in a specific order. Thus, a mobile application can  generally be modeled as a directed acyclic graph (DAG) $G=\{\mathcal{N}, \mathcal{E}\}$, where $\mathcal{N}$ and $\mathcal{E}$ are the sets of nodes (i.e., task modules) and directed edges (i.e., indicating the dependency between two task modules), respectively.   This modeling has been widely adopted in the computation offloading literature, such as, \cite{wu2019efficient,luo2017energy,zhang2015collaborative,mao2017mobile,kao2017hermes}, which  capture the inter-dependency among different task modules using directed acyclic call-graph. In practice, the partition of application into task modules   can be obtained using   task profilers \cite{melendez2017computation,miettinen2010energy} or manually decided by the mobile user.  In Fig. \ref{fig:tag}, we show  an example of a DAG modeled real-world mobile mobile application called Smart Diagnosis.

\begin{table*}[h]
\caption{Frequently used notations.} 
\centering 
\begin{tabular}{c|c | c |c } 
\hline 
 Notations & Descriptions &  Notations & Descriptions\\
 \hline
 $\mathcal{N}$ & set of task modules in an application  &   $c_n\in\{0,1\}$ &  whether or not module $n$ is executed at local device\\
  \hline
 $\mathcal{E}$ & set of dependency connections among   modules & $s_n\in\{0,1\}$ &  whether or not module $n$ is offloaded to edge server\\
  \hline
 $\omega_n$ & workload (in CPU cycles) of task module $n\in\mathcal{N}$ & $Q_{m,n}^{u}$ & uncertain existing  queue  lengths     at  the  client device\\
  \hline
 $o_{n,k}$ & output data size (in bits) from module $n$ and to $k$ & $Q_{m,n}^{d}$ & uncertain existing  queue  lengths     at  the  edge server\\
  \hline
 $\mathcal{C}(n)$ & set of children nodes of module $n$ & $f_c$ ($f_s$) & CPU frequencies at local device (edge server) \\
  \hline
  $\mathcal{M}(n)$ & set of parent nodes of module $n$ & $R_u$  ($R_d$) & uncertain uploading (downloading) bit rates\\
   \hline
  $\Delta$ & duration of a time slot in the MEC system & $P_u$  ($P_d$) & uncertain uploading (downloading) power consumption\\
   \hline
     $T$ & completion deadline of the mobile application & $\epsilon_m$ & upper bound of extreme event occurrence probability\\
   \hline
 $\x_{n,t}^p$ & whether or not  module $n$ terminates   execution in   & $\epsilon$ & approximation factor of the optimal solution\\
 \cline{3-4}
  $\in\{0,1\}$   & $t$th  slot at   client device $p=0$ (or edge   $p=1$) & $\Psi$ & worst-case expected energy consumption of local device\\
\hline
\end{tabular}
\label{freq_notations} 
\end{table*}

A task module $n\in\mathcal{N}$ of a mobile application is defined by $\omega_n$, where $\omega_n$ is the computation workload (in CPU cycles) of module $n$. The dependency between two task modules is defined by $o_{n, k}$, for $k \in \mathcal{C}(n)$, where $o_{n, k}$ is the output data size (in bits) from $n$ to $k$ and $\mathcal{C}(n)$ is the set of child nodes of module $n$. Likewise, we also denote by $\mathcal{M}(n)$ the set of parent nodes of module $n$. Note that, $\omega_n$'s and $o_{n,k}$'s can be inferred by task profilers \cite{melendez2017computation,miettinen2010energy} before running any offloading scheme.   Table \ref{freq_notations} lists the frequently used notations. 

\subsection{MEC System Model}

We consider a typical MEC system, in which task modules of a mobile application are offloaded over uncertain radio channels to edge servers. A task module is waiting in a queue before it is transmitted over a wireless channel. In addition, to support computation offloading, the edge server assigns virtual machines to execute task modules offloaded by client devices. We also take into account a time-slotted MEC system; the time is divided into slots with equal duration $\Delta$, and the discrete time period $((t-1)\Delta, t\Delta]$ is referred to as slot $t$. 
To guarantee the quality of service, the entire mobile application is considered to complete before $T$ timeslots.

\subsection{Computation Execution in the MEC System}

\subsubsection{The Location of Executing A Task Module}

To   model the module execution location, we first define a   binary variable $\x_{n, t}^p$ to indicate whether module $n$ terminates its execution at the $t$th slot at a client device (i.e., $p = 0$) or at an edge server (i.e., $p = 1$). $\x_{n}^0\in\{0,1\}^T$  and $\x_{n}^1\in\{0,1\}^T$ are 1-sparse vectors that recording values of $\x_{n, t}^0$ and $\x_{n, t}^1$, respectively.  As a result, we have $c_n = \sum_{t=1}^T\x_{n, t}^0=1$ and $s_n = \sum_{t=1}^T\x_{n, t}^1=1$ for executing task module $n$ at a client device and an edge server, respectively. Since a task module can be executed at either a client device or the edge server, we have
\begin{equation}
c_n+s_n = 1, \forall n \in\mathcal{N}.
\label{other_node}
\end{equation}
Particularly, for practical mobile applications, the first and last task module are usually used for initializing a mobile application and displaying computation results \cite{mao2017mobile}. Both of them need to be done at the client. Hence, 
\begin{equation}
c_1 = 1 \quad{\rm{and}}  \quad c_N = 1. 
\label{node_1}
\end{equation}

\subsubsection{Module Execution Completion}

The execution completion time of a module $n$ can be calculated by $\sum_{p=0}^1\sum_{t=1}^Tt\x_{n,t}^p$. Due to a time-critical mobile application, the completion time of the last task module is subject to
\begin{equation}
\sum_{t=1}^Tt\x_{N,t}^0\leq T.
\label{deadline}
\end{equation}


\subsubsection{Task Module Execution Dependency}
\label{dependency}

Due to the interaction among task modules, the completion of executing a task module depends on its parent modules. Specifically, the starting time to execute a module $n$ should be no earlier than the time when its parent module $m$ ($\forall m \in \mathcal{M}(n)$) completes   execution plus the time for possible output data queuing and transmission between the client device and the edge server. Thus, we can   model the module execution dependency by
\begin{equation}
\begin{aligned}
\label{dependence}
&\sum_{p=0}^1\sum_{t=1}^Tt\x_{m,t}^p+c_ms_n \frac{Q_{m,n}^u+o_{m,n}}{R_{u}}+s_mc_n\frac{Q_{m,n}^d+o_{m,n}}{R_{d}}\\
&\leq\sum_{p=0}^1\sum_{t=1}^T t\x_{n,t}^p-s_n\frac{\omega_n}{f_s}-c_n\frac{\omega_n}{f_c}, \forall m \in \mathcal{M}(n), \forall n\in\mathcal{N},
\end{aligned}
\end{equation}
where $Q_{m,n}^u$ (resp. $Q_{m,n}^d$) is a random variable presenting the existing queue lengths of the out-going buffers at a client device (resp. edge server) when transmitting $o_{m, n}$ over uncertain wireless channels, $f_c$ and $f_s$ are the CPU frequencies of the client device and virtual machine, respectively, and $R_u$ (resp. $R_d$) is a random variable presenting the bit rates for uploading data to (resp. downloading data from) the edge server.  For instance, $\frac{Q_{m,n}^u+o_{m,n}}{R_{u}}$ calculates the overall time of transmitting the output data of size $o_{m,n}$  and the existing buffered data of size $Q_{m,n}^u$ to the edge server under a random transmission rate $R_u$.  Based on the Shannon-Hartley theorem, the bit rates $R_u$ and $R_d$ can be obtained by $R= B\log_2(1+SNR)$, where $B$ is the bandwidth and $SNR$ is the nondeterministic signal-to-noise ratio. Note that in (\ref{dependence}) we do not have any assumptions on the queuing models (e.g., Poisson distributed task arrival rates and exponential distributed job processing time assumed in the M/M/1 model \cite{ganesh2004big}), and the uncertainties introduced by queue length and bit rates will be addressed by the extreme value theory in the following section.

\section{Handling the Uncertainties}
\label{handlernd}

Due to the uncertain radio channels and network queue length, $Q_{m,n}^u, Q_{m,n}^d, R_{u}$, and $R_{d}$ in (\ref{dependence}) are random variables. To deal with the uncertainties, we propose to apply the extreme value theory and bound the time for the data queuing and transmission in the worst case. In so doing, we can draw broad conclusions about computation offloading in MEC systems without relying on specific case of radio channel and queuing model (e.g., the block-fading channel, constant or Poisson distributed queue size). The reason is that the extreme value theory is a technique describing the unusual rather than the usual, even if the network parameters change over time, their values will still be upper-bounded by their extreme values with a large probability. 
By explicitly accounting for the uncertainties, our methodology is also superior than the classical ones, e.g., \cite{ganesh2004big}, which applies large deviation theory to handle rare events in the network.

Specifically, we consider the Generalized Extreme Value (GEV) distribution instead of its specific cases commonly used by existing works \cite{liu2007application,xiao2012reliable,swapna2013throughput,liu2017latency} when applying the extreme value theory to wireless radio channels. This is because these methods need to choose the most appropriate special case of the GEV distribution, and the followed statistic inferences do not allow any flexibility once the special case is determined.

We start with the case that a client device offloads module $n$ to the edge server, i.e., $s_n=1$, $c_n=0$, and $\x_n^0 = \mathbf{0}$, 
and rewrite (\ref{dependence}) as
\begin{equation}
\begin{aligned}
c_m\frac{Q_{m,n}^u+o_{m,n}}{R_{u}}  \leq \sum_{t=1}^{t=T}t\x_{n,t}^1- \sum_{p=0}^{1}\sum_{t=1}^{t=T} t\x_{m,t}^p-\frac{\omega_n}{f_s},
\end{aligned}
\label{QR}
\end{equation}
$\forall m \in\mathcal{M}(n), \forall n\in\mathcal{N}$. 
Define random variable  $V = \frac{Q_{m,n}^u+o_{m,n}}{R_{u}}\in[0,+\infty)$, which is the quotient  between  two random variables. According to \cite{curtiss1941distribution}, one generic way to derive the distribution of   quotient   from the joint distribution of the two involved  random variables  (i.e., $Q_{m,n}^u+o_{m,n}$ and $R_u$ in our case) is by integration of the following form $f_V(v) = \int |r|f_{Q,R}(vr,r)dr$, where $f_{Q,R}(vr,r)$ is the joint PDF of $Q_{m,n}^u+o_{m,n}$ and $R_{u}$,   $v$ and $r$ denotes one pair of instance of random variable $V$ and $R_u$, and the multiplication of $u$ and $v$ represent one instance of the size of the buffered data ($Q_{m,n}^u$) plus the size of $o_{m,n}$.


The common way of handling the randomness is to bound the expected time of the data transmission and queuing by taking the expected value of $V$. However, due to no prior knowledge of the distributions of $Q_{m,n}^u$ and $R_{u}$, it is impossible to obtain $f_V(v)$ in practice. Moreover, only relying on the expected value is insufficient for ultra-reliable low latency communication (URLLC) applications, as it might fail to quantify some extreme events in the wireless communication systems \cite{liu2017latency,abdel2018ultra}. To address this issue, we consider the $k$th order statistic, i.e.,  $V_{(k)} = \underset{1\leq i \leq k}{\text{max}}V_i$, 
to bound the time of data transmission and queuing in the worst case. $\{V_1,V_2,\cdots V_k\}$ are samples drawn from $f_V(v)$. Thus, we can rewrite (\ref{QR}) as
\begin{equation*}
\begin{aligned}
c_ms_n\V \leq \sum_{t=1}^{t=T}t\x_{n,t}^1-\sum_{p=0,t=1}^{p=1,t=T} t\x_{m,t}^p-s_n\frac{\omega_n}{f_s},
\end{aligned}
\label{Vn}
\end{equation*}
$\forall m \in\mathcal{M}(n), \forall n\in\mathcal{N}$, 
and bound the probability that $V_{(k)}$ exceeds its upper bound by
\begin{equation}
\Pr(c_m\V \geq b_{m,n}^{c,s})\leq \epsilon_m, \forall m \in\mathcal{M}(n), \forall n\in\mathcal{N},
\label{pr}
 \end{equation}
where $b_{m,n}^{c,s} =\sum_{t=1}^{t=T}t\x_{n,t}^1-\sum_{p=0,t=1}^{p=1,t=T} t\x_{m,t}^p-\frac{\omega_n}{f_s}$ and $\epsilon_m\ll1$. To characterize the PDF of $\V$, we resort to the extreme value theory and explore the Generalized Fisher-Tippett-Gnedenko theorem to bound the worst-case data transmission time and queuing delay in the MEC system.

\begin{theorem}[Generalized Fisher-Tippett-Gnedenko Theorem \cite{coles2001introduction}]
\label{theorem_gev}
Let $V_i, 1\leq i \leq k$, be a sequence of random variables with a common PDF as $f_V(v)$, and $\V = \underset{1\leq i \leq k}{\text{max}}V_i$. If there exist two sequences of constants $a_k\in\R^+$ and $b_k\in\R$ such that $\Pr(\frac{\V-b_k}{a_k}\leq z)\approx G(z)$ as $k\rightarrow \infty$
for a non-degenerate distribution function $G$, where $\approx$ stands for ``asymptotic to". Then $G$ is a member of the GEV family with a cumulative distribution function (CDF) as 
\begin{equation*}
  G(z) =\left\{
  \begin{aligned}
    &                               \exp\bigg\{-[1+\xi(\frac{z-\mu}{\sigma})^{-1/\xi}]\bigg\}, & \text{if $\xi \neq 0$,} \\
     &                              \exp\bigg\{ -exp[-\frac{z-\mu}{\sigma}]     \bigg\}, & \text{if $\xi = 0$,}
  \end{aligned}
  \right.
\end{equation*}
defined over the set $\{z:1+\xi(z-\mu)/\sigma>0\}$, where $\mu\in \R$, $\xi\in \R$, and $\sigma\in\R^+$ are the location parameter, shape parameter, and scale parameter, respectively.
\end{theorem}
\begin{remark}\label{remark:cdf_remark} 
In particular, $G(z)$ becomes the  Gumbel distribution, Fr\'echet distribution, and Weibull distribution, when $\xi=0$, $\xi>0$, and $\xi<0$, respectively. The three distributions are all members of GEV family distribution and can be written in the form of $G(z)$ (with different parameters, i.e., $\mu$, $\xi$, and $\sigma$). Theorem \ref{theorem_gev} establishes the CDF in terms of  $\Pr(\frac{\V-b_k}{a_k}\leq z)$ instead of  $\Pr(\V\leq z)$. This is because as $k$ increases the corresponding probability density function  may degenerate to a point mass (page 46 of \cite{coles2001introduction}). It can be avoided by allowing a linear renormalization of the block maximum $V_{(k)}$, i.e., $\frac{V_{(k)}-b_k}{a_k}$, and the renormalized variable is asymptotically distributed as $G(z)$ as $k$ increases. Note that if the renormalization of a variable follows a GEV family distribution, the variable itself also follows a GEV family distribution with different parameters (page 49 of \cite{coles2001introduction}), i.e., the CDF of $\V$ takes the same form    as $\Pr(\frac{\V-b_k}{a_k}\leq z)$ (i.e., $G(z)$), but with different  $\mu$, $\xi$, and $\sigma$. 
\end{remark}

Based on Remark \ref{remark:cdf_remark}, to bound the probability (in (\ref{pr})) by $\epsilon_m$, we can bound the corresponding extreme quantile of $V_{(k)}$, i.e., $z_{\epsilon_m}^u$. 
Then, we have $\Pr(\V\geq z_{\epsilon_m}^u)=\epsilon_m^u$. 
Thus, (\ref{pr}) becomes
\begin{equation}
c_mz_{\epsilon_m}^u\leq b_{m,n}^{c,s}, \forall m \in\mathcal{M}(n),  \forall n \in\mathcal{N}.
\label{extreme_quantiles}
\end{equation}
Note that 
$b_{m,n}^{c,s}$ is determined by the binary variable  $\x_{n}^p$,  and $z_{\epsilon_m}^u$ is a constant given the probability $\epsilon_m$ and a GEV distribution. By inverting the function of $G$ (which can also describe the CDF of $\V$, but with different $\mu$, $\xi$, and $\sigma$), the extreme quantile is 
\begin{equation}
z_{\epsilon_m}^u=\left\{
\begin{aligned}
&\mu-\frac{\sigma}{\xi}[1-\{-\log(1-\epsilon_m)\}^{-\xi}], & \text{if $\xi \neq 0$,}  \\
&\mu-\sigma\log\{-\log(1-\epsilon_m)\}, & \text{if $\xi = 0$.}
\end{aligned}
\right.
\label{z}
\end{equation}
Parameters of the GEV distribution $\mathbf{\Theta} = (\xi,\sigma,\mu)$ can be inferred via the maximum likelihood estimation (MLE), which will be elaborated in  Section \ref{inf}.

Now, we consider   $n$ is executed at the client device, i.e.,   $c_n=1, s_n=0$ and $\x_n^1=\mathbf{0}$. We can rewrite (\ref{dependence}) as 
\begin{equation*}
\begin{aligned}
s_m\frac{Q_{m,n}^d+o_{m,n}}{R_{d}}  \leq \sum_{t=1}^{t=T}t\x_{n,t}^0-\sum_{p=0,t=1}^{p=1,t=T} t\x_{m,t}^p-c_n\frac{\omega_n}{f_c},
\label{sc}
\end{aligned}
\end{equation*}
$\forall m \in\mathcal{M}(n), \forall n\in\mathcal{N}$. 
We notice that $\frac{Q_{m,n}^d+o_{m,n}}{R_{d}}$ is also a random variable. Thus, we also apply the $k$th order statistics and the extreme value theory to obtain
\begin{equation}
s_mz_{\epsilon_m}^d \leq b_{m,n}^{s,c},  \forall m \in\mathcal{M}(n), \forall n\in\mathcal{N}.
\label{extreme_s}
\end{equation}
where $z_{\epsilon_m}^d$ is the extreme quantile and $b_{m,n}^{s,c} = \sum_{t=1}^{t=T}t\x_{n,t}^0-\sum_{p=0,t=1}^{p=1,t=T} t\x_{m,t}^p-\frac{\omega_n}{f_c}$ is the nondeterministic upper bound.

\section{Problem Formulation}
\label{PF}

We consider to minimize a client device's energy consumption under the constraint of the completion time of the mobile application. Specifically, a client device's energy consumption for task execution and data transmission can be given by
\begin{equation}
\begin{aligned}
E = \sum_nc_n\kappa \omega_nf_c^2 &+ P_{u}\sum_{m\in\mathcal{M}(n)}\sum_n   c_ms_n\frac{o_{m,n}}{R_{u}}\\
&+P_{d}\sum_{m\in\mathcal{M}(n)}\sum_ns_mc_n\frac{o_{m,n}}{R_{d}},
\end{aligned}
\label{energu}
\end{equation}
where $\kappa$ is a constant related to the hardware architecture of the client device \cite{mao2017mobile,miettinen2010energy},  $P_{u}$ and $P_{d}$ are random variables represent the power consumption of the device for transmitting data to and receiving data from the server. Due to the uncertain radio channels, the client device's energy consumption is nondeterministic as well. 

To design a robust computation offloading scheme for transmitting offloaded computation over uncertain radio channels, we first formulate the energy consumption in the worst case: 
\begin{equation}
\begin{aligned}
\bar{E} = \sum_nc_n\kappa \omega_nf_c^2 &+ \sum_{m\in\mathcal{M}(n)}\sum_n   c_ms_n o_{m,n} J_{(k)}\\
&+\sum_{m\in\mathcal{M}(n)}\sum_ns_mc_n o_{m,n}H_{(k)},
\end{aligned}
\label{energy_worst}
\end{equation}
where $J_{(k)} = \underset{1\leq i \leq k}{\text{max}}J_i$ and $H_{(k)} = \underset{1\leq i \leq k}{\text{max}}H_i$ are the $k$-th order statistics. 
$\{J_1,J_2,\cdots, J_k\}$ and $\{H_1,H_2,\cdots, H_k\}$ are the sample sets drawn from $f_{P_u/R_u}(p_u/r_u)$ and $f_{P_d/R_d}(p_d/r_d)$, respectively (by following statistical conventions, $P_u/R_u$ and $P_d/R_d$ denote the random variables, and $p_u/r_u$ and $p_d/r_d$ are  the instances). 
Then, we define the expected energy consumption of the client device in the worst case as follows:
\begin{equation}
\begin{aligned}
\Psi  =&\qquad\mathbb{E}_{\mathclap{\substack{\qquad\\\qquad\\J_{(k)}\sim GEV(j_{(k)})\\ H_{(k)}\sim GEV(h_{(k)})}}} \quad(\bar{E})  \sum_nc_n\kappa \omega_nf_c^2
+\sum_{m\in\mathcal{M}(n)}\sum_n\\ 
&\Big( c_ms_no_{m,n}\quad\E_{\mathclap{\substack{\qquad\\\qquad\\J_{(k)}\sim GEV(j_{(k)})}}}(J_{(k)})+ s_mc_no_{m,n}\quad \E_{\mathclap{\substack{\qquad\\\qquad\\H_{(k)}\sim GEV(h_{(k)})}}}(H_{(k)})\ \ \Big),
\end{aligned}
\label{energy_worst_ex}
\end{equation}
where $GEV(j_{(k)})$ and $GEV(h_{(k)})$ are the inferred probability distributions of $J_{(k)}$ and $H_{(k)}$, respectively. 
The expected value of GEV distributed variable can be calculated as \cite{coles2001introduction}
\begin{equation}
\E_{\mathclap{\substack{\qquad\\\qquad\\X\sim GEV(x)}}}(X) = 
\begin{cases}
\mu+\sigma(g_1-1)/\xi, & \text{if $\xi \neq 0$, $\xi <1$}, \\
\mu+\sigma\gamma, & \text{if $\xi = 0$},\\
\infty, & \text{if $\xi \geq 1$},
\end{cases}
\label{mean}
\end{equation}
where $g_1 = \Gamma(1-\xi)$ and $\gamma$ is Euler's constant. For notational simplicity, we have $\theta_u = \quad\mathbb{E}_{\mathclap{\substack{\qquad\\\qquad\\J_{(k)}\sim GEV(j_{(k)})}}}(J_{(k)})$ and $\theta_d = \quad\E_{\mathclap{\substack{\qquad\\\qquad\\H_{(k)}\sim GEV(h_{(k)})}}}(H_{(k)})$.


Therefore, we formulate the energy efficient computation offloading problem as
\begin{equation}
\begin{aligned}
\text{MP}:\qquad& \underset{\x_{n}^p\in\{0,1\}^T,\ p\in\{0,1\}}{\text{min}} \quad \Psi\\
&\text{\textbf{s. t.}}\quad (\ref{other_node}),(\ref{node_1}),(\ref{deadline}),(\ref{extreme_quantiles}),(\ref{extreme_s}).
\label{formulation}
\end{aligned}
\end{equation}

\section{A Column Generation Based Efficient $\epsilon$-bounded Approximation Algorithm}
\label{CG}

The formulated optimization problem in (\ref{formulation}) is a quadratically constrained binary quadratic programming. In particular, the decision variables are $2N$ 1-sparse binary vectors in $\{0,1\}^T$, resulting in a total of $2^{2NT}$ degrees of freedom. This makes the formulated problem computationally-intractable. To address this issue, we propose an $\epsilon$-bounded approximate algorithm based on the column generation (CG) technique that is usually used to solve linear or nonlinear programming in an iterative way \cite{bertsimas1997introduction} and is widely used in wireless communication \cite{li2014energy,li2014optimal}. Particularly, our proposed scheme can find the $\epsilon$-bounded approximate offloading policy, and especially the optimal policy when setting $\epsilon= 0$.

Specifically, we consider MP in (\ref{formulation}) as a master problem, and solve it via starting with a restricted master problem (i.e., called RMP) which minimizes $\Psi$ given only a fraction of the columns. We can add a new column into RMP only if it has been determined by a price problem (i.e., called PP) to be profitable for further reducing $\Psi$. To be more specific, at each iteration, PP first determines whether any column ($\x_n^p$) uninvolved in RMP can lead to a negative reduced cost, and then the one with the most negative reduced cost will be added into RMP. The iteration terminates at or satisfyingly close to the optimal solution. 
\subsection{Restricted Master Problem}
The general RMP that minimizes $\Psi$ by considering only a subset of all decision columns is 
\begin{equation}
\begin{aligned}
\text{RMP}:\qquad& \underset{\x_n^p\in\mathcal{S}_I}{\text{min}}\quad\widetilde{\Psi}\\
&\ \  \text{\textbf{s. t.}}\quad (\ref{other_node}), (\ref{node_1}), (\ref{deadline}), (\ref{extreme_quantiles}), (\ref{extreme_s}),
\label{rmp}
\end{aligned}
\end{equation} 
where $\mathcal{S}_I$ is a subset of all columns considered in MP, denoted as $\mathcal{S} = \{\x_1^0\}\cup\{\x_N^0\}\cup\big\{\x_n^p|\x_n^p\in\{0,1\}^T, n\in\{2,\cdots,N-1\}, p\in\{0,1\}\big\}$, and $\widetilde{\Psi}$ is the  expected worst case energy consumption calculated with respect to $\mathcal{S}_I$. At the same time, we set all the elements in set $\mathcal{S}_I^c = \mathcal{S}\backslash\mathcal{S}_I$ (the complement set of $\mathcal{S}_I$) to be zero vectors. Given a feasible offloading policy determined by $\big\{\x_n^p, n\in\mathcal{N}, p\in\{0,1\}\big\}$, $\widetilde{\Psi}$ becomes deterministic. Thus, RMP is reduced to a feasibility checking problem. Specifically, with a provided offloading policy, we need to check whether there exists at least one feasible solution in the set defined by the constraints in (\ref{rmp}).

Without loss of generality (w.l.o.g.), we can start with a special RMP in which all computation task modules are assigned to the client device ($c_n=1, s_n=0$,   $\forall n\in\mathcal{N}$). As a result, $\mathcal{S}_I = \big\{\x_n^0|\forall n\in\mathcal{N}\big\}$ and $\widetilde{\Psi} = \sum_n\kappa \omega_nf_c^2$. Then, one feasible solution for this special RMP is $\big\{\x_{n,t}^0=1, {\mathrm{if}}\ t = \sum_{i=1}^n\ceil{\frac{\omega_n}{f_c}},\x_n^1=\mathbf{0}|n\in\mathcal{N}\big\}$. 


Since RMP involves only a subset of $\mathcal{S}$ used by MP, i.e., $\mathcal{S}_I\subset\mathcal{S}$, $\widetilde{\Psi}$ serves as an upper bound on the optimal result of MP \cite{bertsimas1997introduction}. By introducing more columns to RMP, the upper bound may be lowered down further. Therefore, we need to determine which column is the most profitable to MP in terms of negative reduced cost, and when the optimal result of RMP is exactly the same as or satisfyingly close to the optimal result of MP. This is achieved by solving the corresponding price problem, described in the following.

\subsection{Adding A New Column to RMP via PP}

In each iteration, when RMP is solved, we need to check whether adding any new column can lead to better solution to MP. In particular, for each $\x_n^1$ in  $\mathcal{S}_I^c$ (i.e.,  $\mathcal{S}\backslash\mathcal{S}_I$), 
we need to examine whether it has a negative reduced cost. The reduced cost $\zeta_n$ for a column $\x_n^1\in\mathcal{S}_I^c$ is,  according to \cite{bertsimas1997introduction},   
\begin{equation}
\begin{aligned}
\zeta_n=&\sum_{m\in\mathcal{M}(n)}   c_ms_no_{m,n}\theta_u+\sum_{k\in\mathcal{C}(n)} s_nc_ko_{n,k} \theta_d\\   
&-\sum_{m\in\mathcal{M}(n)}\pi_{m,n} b_{m,n}^{c,s}-\sum_{k\in\mathcal{C}(n)}\pi_{n,k} b_{n,k}^{s,c},
\label{pp}
\end{aligned}
\end{equation}
where $\pi_{m,n}$'s and $\pi_{n,k}$'s are the Lagrangian dual optimal solution corresponding to constraints (\ref{extreme_quantiles}) and (\ref{extreme_s}) in (\ref{rmp}).

Then, we need to find the column that can produce the most negative reduced cost. Hence, the column to be generated and added to RMP is obtained by solving the   price problem:
\begin{equation}
\begin{aligned}
\text{PP}:\qquad& \underset{\x_n^1\in\mathcal{S}_I^c}{\text{min}}\quad\zeta_n\\
&\ \  \text{\textbf{s. t.}}\ \ \   c_ms_nz_{\epsilon_m}^u\leq b_{m,n}^{c,s}, \forall m \in\mathcal{M}(n),\\
&\ \quad\quad\ \ \  s_nc_kz_{\epsilon_m}^d \leq b_{n,k}^{s,c},\ \  \forall k \in\mathcal{C}(n).
\label{pp2}
\end{aligned}
\end{equation}
In PP, the new column $\x_n^1$ is embedded in $b_{m,n}^{c,s}$ and $b_{n,k}^{s,c}$. This is because all execution location indicators $c_m$'s, $s_n$'s and $c_k$'s have been determined by solving RMP. Thus, PP is a binary integer programming (BIP). Denote by $r^*$ the optimal solution to PP. If $r^*\geq 0$, then no column can generate negative reduced cost, and the current solution to RMP is the optimal solution to MP. Otherwise we add to RMP the column with the most negative reduced cost identified by (\ref{pp2}). 

We first discuss the feasible solution to RMP with the added new column and solve PP in the next subsection. Recall that the objective function in (\ref{rmp}) becomes deterministic under a new  offloading policy. The new feasible solution can be obtained by setting
\begin{equation}
\x_g^0 = \mathbf{0}\quad{\rm{and}}\quad \x_{g,t}^1 = 1,\ \exists_{=1} t\in\{t_{min}, \cdots, t_{max}\},
\label{rmp_f}
\end{equation}
where $g$ is the node index of the new column identified by (\ref{pp2}), $\exists_{=1}$ means ``exists and choose one'', $t_{min} = \underset{m\in\mathcal{M}(n)}{\text{max}}\sum_{p=0}^{p=1}\sum_{t=1}^{t=T}t\x_{m,t}^p+c_ms_no_{m,n}\theta_u+\ceil{\frac{\omega_n}{f_s}}$ and $t_{max} = \underset{k\in\mathcal{C}(n)}{\text{min}}\sum_{p=0}^{p=1}\sum_{t=1}^{t=T}t\x_{k,t}^p-s_nc_ko_{n,k}\theta_d-\ceil{\frac{\omega_k}{f_c}}$ are constants, indicating the range of the execution completion time of node $g$ at the server. $t_{min}\leq t_{max}$ is guaranteed when solving PP.

\subsection{Solving PP}
\label{PP_solve}

To solve PP, we need to evaluate the reduced cost of every $\x_n^1\in \mathcal{S}_I^c$ by solving $N-2$ BIP's and checking their sign. Yet, solving multiple BIP's is time-consuming when a DAG has hundreds or thousands of nodes. To circumvent this problem, we formulate a new price problem (NPP): 
\begin{equation}
\begin{aligned}
\text{NPP}:\quad& \underset{\lambda_n,\ \x_n^1,\ n\in\mathcal{S}_I^c}{\text{min}}\ \ \ \ \ r  =\sum_{n\in\mathcal{S}_I^c}\lambda_n\zeta_n\\
&\text{\textbf{s. t.}}\quad\   c_ms_nz_{\epsilon_m}^u\leq \sum_{n\in\mathcal{S}_I^c}\lambda_nb_{m,n}^{c,s}, \forall m \in\mathcal{M}(n),\\
& \quad \quad\quad\ s_nc_kz_{\epsilon_m}^d \leq \sum_{n\in\mathcal{S}_I^c}\lambda_nb_{n,k}^{s,c},\ \  \forall k \in\mathcal{C}(n),\\
& \quad \qquad\sum_{n\in\mathcal{S}_I^c}\lambda_n = 1,\ \ \ \lambda_n\in\{0,1\},
\label{pp3}
\end{aligned}
\end{equation}
where the decision variables are $\lambda_n$'s and $\x_n^1$'s (absorbed in $b_{m,n}^{c,s}$, $b_{n,k}^{s,c}$ and $\zeta_n$). (\ref{pp3}) is equivalent to (\ref{pp2}), because in (\ref{pp3}) only a single $\lambda_n$ can be activated (equals to 1) in both the objective function and the constraints, and the number of $\lambda_n$'s equals to the cardinality of $\mathcal{S}_I^c$ in (\ref{pp2}). Although NPP formulated in (\ref{pp3}) is a quadratic integer programming with quadratic constraints, which is generally more complicated than (\ref{pp2}), we can solve it efficiently by further decomposing (\ref{pp3}) into two separate problems, including a column selection (CS) problem with the fixed $b_{m,n}^{c,s}$ and $b_{n,k}^{s,c}$ and a completion time decision (TD) problem with the selected columns.

W.l.o.g., we can initially  set $b_{m,n}^{c,s} = z_{\epsilon_m}^u$ and $b_{n,k}^{s,c} = z_{\epsilon_m}^d$, and optimize (\ref{pp3}) over $\lambda_n$ as
\begin{equation}
\begin{aligned}
\text{CS}:\qquad& \underset{\lambda_n,\ n\in\mathcal{S}_I^c}{\text{min}}\ \ \ \ \ r  =\sum_{n\in\mathcal{S}_I^c}\lambda_n\zeta_n\\
&\text{\textbf{s. t.}}\quad\  \sum_{n\in\mathcal{S}_I^c}\lambda_n = 1,\ \ \ \lambda_n\in\{0,1\}.
\label{CS}
\end{aligned}
\end{equation}
Then (\ref{pp3}) is reduced to a trivial problem, i.e., $0$-$1$ knapsack problem with the maximum weight capacity $W=1$. The solution to CS is $\lambda_l=1$, for $l = \underset{n}{\text{arg\ min}}\ \zeta_n$.

Next, we fix $\lambda_n$'s, and formulate the TD problem as follows:
\begin{equation}
\begin{aligned}
\text{TD}:\qquad& \underset{\x_l^1,\ l\in\mathcal{S}_I^c}{\text{min}}\ \ \ \ \ r  =\lambda_l\zeta_l\\
& \text{\textbf{s. t.}}\qquad  c_ms_lz_{\epsilon_m}^u\leq \lambda_lb_{m,l}^{c,s}, \forall m \in\mathcal{M}(l),\\
&  \qquad\quad\ \  s_lc_kz_{\epsilon_m}^d \leq \lambda_lb_{l,k}^{s,c},\ \  \forall k \in\mathcal{C}(l),\\
&\qquad\quad\ \  \lambda_l=1.
\label{TD}
\end{aligned}
\end{equation}
TD is a BIP problem with the only decision variable $\x_l^1$ (absorbed in $b_{m,l}^{c,s}$, $b_{l,k}^{s,c}$ and $\zeta_l$). 
By relaxing $\x_l^1$ as $0\leq \x_{l,t}^1\leq 1$, TD can be relaxed to a linear programming that can be easily solved by polynomial interior algorithm \cite{bertsimas1997introduction}. Then, we set one element with the largest value in the solution vector as $\x_{l,t}^1=1$, and other elements as $\x_{l,t'}^1=0$ ($t'\neq t$).

We use $\underline{r}^*$ to denote the solution of NPP achieved by solving CS and relaxed TD iteratively, and summarize the above process of solving NPP in Algorithm~\ref{solvepp}.
\begin{algorithm}
\caption{Solving NPP}
\begin{flushleft}
\textbf{Input:} Dual solution $\pi_{m,n}$'s and $\pi_{n,k}$'s of RMP, $\mathcal{S}_I^c$, maximum iteration number $max\_iter\_num$,  $iter\_num=0$\\
\textbf{Output:} $\x_g^1\in \mathcal{S}_I^c$ (column to be added to RMP)
\end{flushleft}
\begin{algorithmic}[1]
		\While {$iter\_num<max\_iter\_num$ or $\underline{r}^*$ changes}
		\State Initialize $b_{m,n}^{c,s}$'s as $z_{\epsilon_m}^u$, $b_{n,k}^{s,c}$ as $z_{\epsilon_m}^d$;
		\State {Given $b_{m,n}^{c,s}$'s and $b_{n,k}^{s,c}$'s, solve CS by setting $\lambda_l=1$, for $l = \underset{n}{\text{argmin}}\ \zeta_n$};
		\State Given $\lambda_l=1$, relax $0\leq \x_{l,t}^1\leq 1$ and solve the relaxed TD by polynomial interior algorithm;
		\State Set the maximum $\x_{l,t}^1$ to $1$ and other $\x_{l,t}^1$'s as $0$;
				\State Evaluate $\underline{r}^*$;
		\State $iter\_num = iter\_num+1$;
\EndWhile
\State set $g=l$, return $\x_g^1$.
\end{algorithmic}
\label{solvepp}
\end{algorithm}

\subsection{$\epsilon$-Bounded Approximate Solution}

The solver developed in Algorithm~\ref{solvepp} does not achieve the optimal solution to PP, i.e., $r^*$. However, when the result of NPP, i.e., $\underline{r}^*$, is larger than or equal to $0$, i.e., $r^*\geq \underline{r}^*\geq 0$, MP can still be solved optimally.  Even if $\underline{r}^*<0$, we can still find an $\epsilon$-bounded approximate solution to MP. First, we define the $\epsilon$-bounded approximate solution as follows.
\begin{mydef}
The $\epsilon$-Bounded Approximate Solution.  Let $0\leq \epsilon <1$ be the predefined parameter and $\Psi^*$ be the optimal result. Then a solution $\Psi$ is called the $\epsilon$-bounded approximate solution if it satisfies $\Psi^*\leq \Psi \leq (1+\epsilon)\Psi^*$.
\label{def}
\end{mydef}
We can have the following lemma on determining the $\epsilon$-bounded approximate offloading policy.
\begin{lemma}\label{lemma:condition}
Denote by $\Psi_l$ and $\Psi_u$ the lower and upper bounds on the optimal result $\Psi^*$ of MP, respectively. Then, the $\epsilon$-bounded approximate solution can be obtained when no new column ($\x_n^1$) can be found by solving PP, or the iteration stops at $\underline{r}^*\geq 0$, or $\frac{\Psi_l}{\Psi_u}\geq \frac{1}{1+\epsilon}$.
\end{lemma}
\begin{IEEEproof}[Proof]
When $\frac{\Psi_l}{\Psi_u}\geq \frac{1}{1+\epsilon}$, we   have $\Psi_u\leq (1+\epsilon) \Psi_l\leq (1+\epsilon)\Psi^*$. Then, given any result $\Psi$ between the lower and upper bounds, i.e., $\Psi_l\leq \Psi \leq \Psi_u$, we can have $\Psi^*\leq\Psi\leq\Psi_u\leq(1+\epsilon)\Psi^*$, which is the $\epsilon$-bounded approximate solution by  Definition \ref{def}. Besides, when $\underline{r}^*\geq0$ or no new column can be found by solving PP, as discussed before, the obtained solution is the optimal solution and hence an $\epsilon$-bounded approximate solution as well ($\epsilon=0$).
\end{IEEEproof}

The optimal result of RMP at each iteration serves as an upper bound of MP, i.e., $\Psi_u$. A lower bound can be obtained by setting $\Psi_l  =\Psi_u+\mathcal{K}r^*\leq \Psi^*$ according to \cite{bertsimas1997introduction}, where $r^*$ is the optimal solution to PP, and $\mathcal{K}\geq \sum_{n}s_n$ holds for the optimal solution to RMP. Since we do not actually obtain $r^*$ with the decomposition of NPP into CS and relaxed TD, the lower bound can be set to $\Psi_u + \mathcal{K}\underline{r}^*$, which is less than $\Psi_u+\mathcal{K}r^*$ and hence $\Psi^*$. Additionally, because $r^*$ is negative, $\Psi_l$ may be negative as well. Thus, in practice, we set $\Psi_l = {\rm{max}}\{0, \Psi_u+\mathcal{K}\underline{r}^*\}$. 

At each iteration, by solving RMP and NPP we can obtain a new pair of lower and upper bounds. By evaluating their ratio, i.e., $\frac{\Psi_l}{\Psi_u}$, we can determine whether we have obtained an $\epsilon$-bounded approximate solution. The process to find an $\epsilon$-bounded approximate offloading solution for offloading a DAG modeled application is summarized in Algorithm~\ref{offload}.
\begin{algorithm}
\caption{An $\epsilon$-bounded approximate solution for DAG-modeled application offloading}
\begin{flushleft}
\textbf{Input:} initialize $\epsilon$, $\mathcal{S}_I$, $\Psi_l=0$, $\Psi_u=\infty$, $\underline{r}^* = \infty$\\
\textbf{Output:} offloading decision $c_n$'s and $s_n$'s
\end{flushleft}
\begin{algorithmic}[1]
		\While {PP generates a new column, or $\frac{\Psi_l}{\Psi_u}<\frac{1}{1+\epsilon}$, or $\underline{r}^*<0$}
		\State Solve RMP under current $\mathcal{S}_I$ by checking the feasibility, obtain $\Psi_u$ and the dual optimal solutions $\pi_{m,n}$'s and $\pi_{n,k}$'s;
				\State Solve NPP using Algorithm~\ref{solvepp}, and obtain new column $\x_g^1$ according to (\ref{rmp_f});
		\State Obtain the result of NPP, i.e., $\underline{r}^*$;
		\State Update $\mathcal{S}_I  = \mathcal{S}_I\cup\x_g^1$, set $\x_g^0 = \mathbf{0}$, $\mathcal{S}_I^c=\mathcal{S}\backslash\mathcal{S}_I$;
		\State $\Psi_l = \Psi_u+\mathcal{K}\underline{r}^*$;
		\EndWhile
		      \For{\texttt{$n\in\{1,\cdots,N\}$}}
        \State $c_n = \sum_t\x_{n,t}^0$; $s_n = \sum_t\x_{n,t}^1$;
      \EndFor
\end{algorithmic}
\label{offload}
\end{algorithm}

\subsection{Computational Complexity Analysis}
It is obvious that the computational complexity is closely related to the DAG's structure, and the number of parent modules of a node is clearly upper bounded by $N$. 
Since Algorithm \ref{offload} solves the formulated problem iteratively, we assume that it solves the problem by $K$ iterations. There are a total of $N-2$ task modules that can be offloaded, so $K\leq N-2$. At each iteration, we need to solve both the RMP and NPP. As a result, we can analyze the computational complexity as follows. First, solving the RMP is reduced to feasibility checking, which requires the computational complexity of $O(NT)$. Second, the NPP includes CS and TD. Solving CS takes the $\mathcal{O}(N|\mathcal{S}_I^C|)$ computational complexity, where $|\mathcal{S}_I^C|$ decreases by one at each iteration. Solving the TD by the polynomial interior algorithm requires the computational complexity of $\mathcal{O}(n^3)$, where $n$ is the number of variables. Moreover, $\x_l^1$ with $T$ dimensions is the only decision variable, which means we have $n = T$. Therefore, we can conclude that the computational complexity of solving MP in (\ref{formulation}) by the proposed scheme based on column generation is $\mathcal{O}\big(K(NT+N^2+T^3)\big)$. Computational complexity also implicitly 
depends on   $\epsilon$ through the parameter $K$. The reason is that   $K$ depends on 
  if $\frac{\Psi_l}{\Psi_u}<\frac{1}{1+\epsilon}$ holds  (i.e., one of the condition on obtaining an $\epsilon$-bounded approximate solution discussed in Lemma \ref{lemma:condition}). As a result, by choosing different values of $\epsilon$, the value of $K$ also varies.

\subsection{Offloading   Mobile Applications with Special Structures}
Some mobile applications may have simple structures, and can be modeled as DAGs with only sequential or parallel task module dependency \cite{zhang2015collaborative,ra2011odessa}. In this section, we provide principles to conduct computation offloading for these special mobile applications.

\begin{figure}[htb]
  \begin{center}
   \begin{tabular}{cc}
     \includegraphics[width= .4\columnwidth]{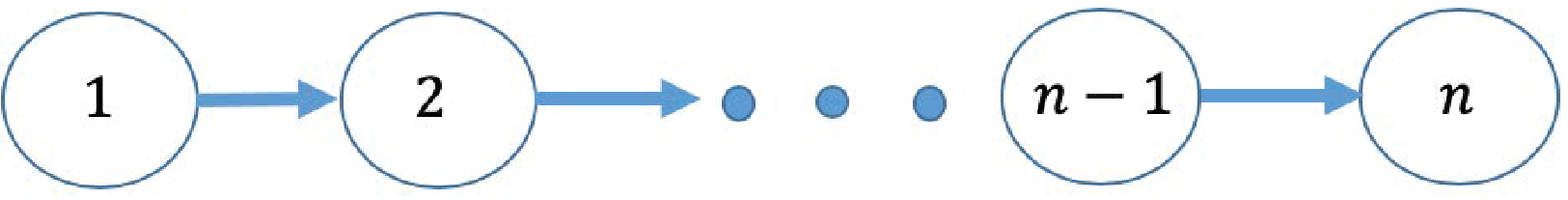}
		&
     \includegraphics[width= .4\columnwidth]{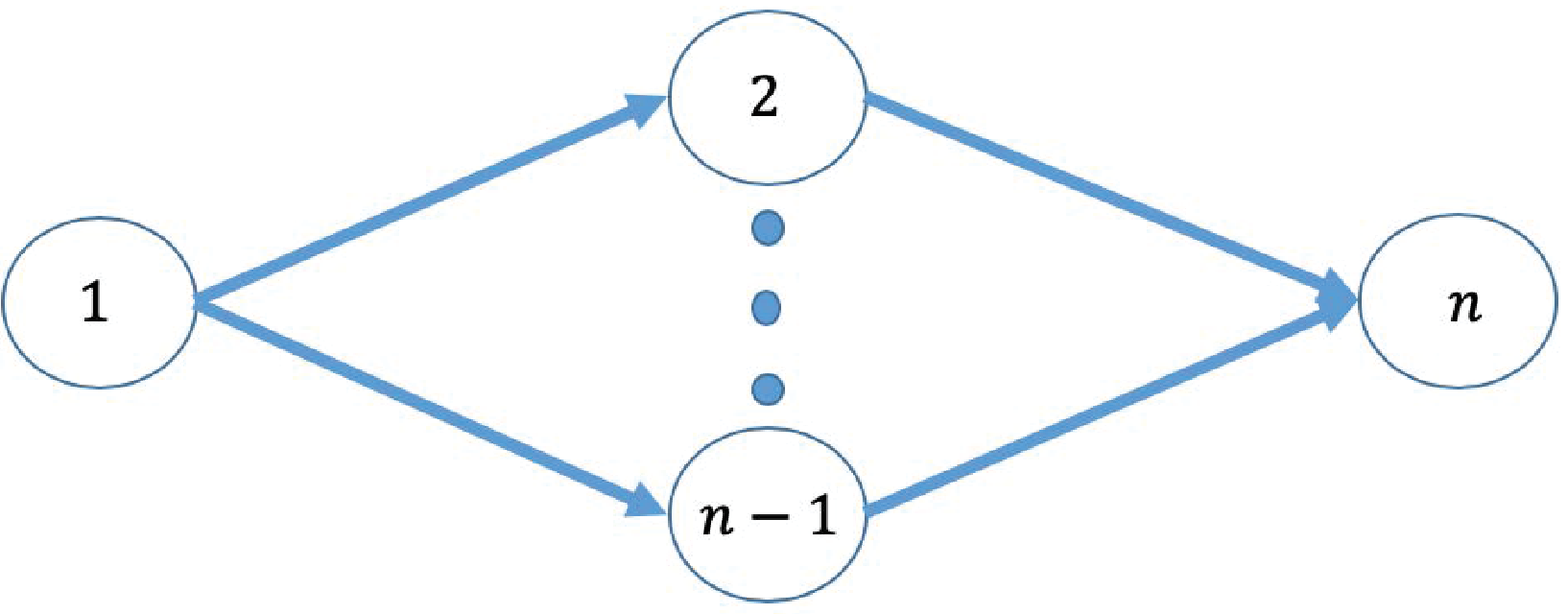}
         \\
    {\small  (a) Sequential dependency}  &
    {\small  (b) Parallel dependency}
        \end{tabular}
       \end{center}
  \caption{\label{fig:special} DAGs with only sequential or parallel task module dependency.}
\end{figure}

\subsubsection{Case $\uppercase\expandafter{\romannumeral1}$: Applications with Only Sequential Module Dependency}
For applications that can be modeled as a DAG with only sequential dependency, i.e., Fig. \ref{fig:special}(a), we develop an efficient way to offload its computations. First, we arrive at a theorem about the optimal computation offloading policy for executing applications with only sequential dependency under uncertain radio channel and network queues.

\begin{theorem}
Under uncertain radio channel with queuing delay, the optimal offloading policy for executing applications with only sequential dependency 
only migrates computations once from the client device to the edge server if needed.
\label{seq_dep}
\end{theorem}
\begin{IEEEproof}[Proof]
W.l.o.g., suppose an application has $N$ sequential modules, and two subsequences, i.e., modules from $u$ to $v$ and $p$ to $q$, are migrated to the server for execution, where $1<u<v<p<q<N$. Then, the expected worst-case energy consumption for this offloading policy is $\Psi_1  =\sum_{n\in S_1}  c_n\kappa \omega_nf_c^2+P_u(o_{u-1}+o_{p-1})\theta_u+P_d(o_{v+1}+o_{q+1})\theta_d$,
where $S_1 = \big\{\{2,\cdots, u-1\} \cup \{v+1,\cdots, p-1\}   \cup \{q+1,\cdots, N-1\}  \big\}$. If modules from $v+1$ to $p-1$ are executed at the server instead, then the expected worst-case energy consumption for the new policy is $\Psi_2  =\sum_{n\in S_2}  c_n\kappa i_n\omega_nf_c^2+P_uo_{u-1}\theta_u+P_do_{q+1}\theta_d$,
where $S_2 = \big\{\{2,\cdots, u-1\} \cup \{q+1,\cdots, N-1\}  \big\}$. Obviously, we have $\Psi_2<\Psi_1$, thus offloading only one subsequence, i.e., migrating only once, provides a better solution.
\end{IEEEproof}

With Theorem \ref{seq_dep}, we can have a   more efficient way to solve PP for applications with only sequential dependency. Since the offloaded modules should be consecutive in a DAG with only sequential dependency, once we find one column with negative reduced cost, we can keep adding its neighboring module with increasing hops as the new column until the reduced cost is positive. Then the subsequence  (each node has a negative reduced cost) is migrated to the server for execution. 

\subsubsection{Case $\uppercase\expandafter{\romannumeral2}$: Applications with Only Parallel Dependency}
We now consider DAG with only parallel dependency, i.e., Fig. \ref{fig:special}(b), 
We also have a theorem on the computation offloading policy for executing such applications. 
\begin{theorem}
Under uncertain radio channel with queuing delay, the optimal offloading policy for executing applications with only parallel dependency is the one that offloads all modules whose energy consumption for execution is larger than the expected worst-case energy consumption for data transmission, i.e., $\kappa \omega_nf_c^2>P_u\theta_uo_1+P_d\theta_do_n, n\in\mathcal{N}$.
\label{para_dep}
\end{theorem}
\begin{IEEEproof}[Proof]
Suppose the minimum expected worst-case energy consumption without the $n$th module 
is $\Psi^*_{\mathcal{N}/n}$. If $n$ is executed locally, then $\Psi  = \Psi^*_{\mathcal{N}/n}+\kappa \omega_nf_c^2$, else if $n$ is executed remotely, then $\Psi = \Psi^*_{\mathcal{N}/n}+P_u\theta_uo_1+P_d\theta_do_n$. Hence, offloading a module that consumes higher energy by local execution than data transmission will result in lower $\Psi$.
\end{IEEEproof}

\section{Experimental Results}
\label{ex}

Now, we will  describe the experiment setup, discuss parameter estimation for the considered GEV distributions, evaluate the performance of the proposed computation offloading scheme, empirically corroborate the  performance degradation caused by simple assumptions on network condition and queues, and compare with the state-of-the-art mechanism. Furthermore, we will also study the scalability of the proposed scheme, and investigate the offloading performance on simulated applications with sequential or parallel dependence.

\subsection{Experimental Setup}
\label{expsetup}

\begin{figure}[h]
    \centering
    \includegraphics[width=.3 \textwidth]{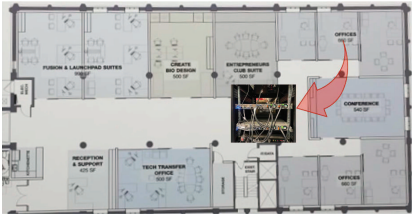}
    \caption{Floor plan of the experiment environment. The edge server is located at the second office room (indicated by the red arrow), top right corner.} 
    \label{fig:floorplan}
\end{figure}

We conduct extensive experiments including both testbed experiments (Section  \ref{exp:performance}, \ref {exp:performance_degrade}, and \ref{exp:compare}) and simulations (Section  \ref{exp:scale} and  \ref{exp:special}) to evaluate the performance of our proposed scheme. Specifically, we establish a MEC architecture in which four servers (provided by the Accipiter System\footnote{\url{https://www.accipitersystems.com/}}), each having $2.4$GHz CPU with 6 cores, 
are deployed in a $40\times 70m^2$ floor area in an office building. The servers are directly connected to a switch. The client devices used in the experiments are all Motorola moto $G^4$ with $1.5$GHz CPU. 
In Fig. \ref{fig:floorplan}, we provide the floor plan of the  environment along with the edge server used in the testbed experiments. 

We execute a real-world mobile application, i.e., Smart Diagnosis visualized in Fig. \ref{fig:tag}. We implement this application on Android devices and encapsulate the proposed  scheme as an API that can be called when the application is invoked. We set the competition time   $T=0.5\times 10^4$ milliseconds (approximately the average execution time of a mobile device running a computer vision   application), and 
$\Delta=1$ millisecond.

To   measure the communication parameters in the MEC system, we first execute Smart Diagnosis locally on several mobile devices, randomly walk in the considered area 
with varying speeds, make video phone calls, and run programs at these devices to generate random matrices with varying sizes,  offload them to the servers to calculate the singular values, then receive the results. At the same time, we also use available Android tools, i.e., tPacketCapture
,
WiFi SNR
,
and PETrA \cite{di2017software}, to record the statistics of the queue length, the channel SNR, and the client device's energy consumption, respectively. Specifically, tPacketCapture captures out-going and incoming packets using VpnService provided by Android OS, and the captured data are saved as PCAP files 
and further analyzed using Wireshark. WiFi SNR records the varying SNR, from which we calculate the corresponding bit rates using Shannon-Hartley theorem given the predetermined bandwidth for data transmission. PETrA provides the client device's energy consumption profile in real-time when executing applications. In Table \ref{table:samle_statistics}, we show some statistics of the out-going and incoming packet size (i.e., $P_u$ and $P_d$), uploading and downloading data rates (i.e., $R_u$ and $R_d$), and uploading and downloading power consumption (i.e., $P_u$ and $P_d$).


\begin{table}[h]
\caption{Statistics  of collected samples.} 
\centering 
\begin{tabular}{c|c| c} 
\hline
parameters& mean & standard deviation\\
\hline 
$P_u$ ($\times 10^5$ bit) & $1.01$ & 0.09\\
$P_d$ ($\times 10^5$ bit) &1.32   & $0.01$ \\
$R_u$ ($\times 10^5$ bit/s) & 4.5  &   0.5\\
$R_d$ ($\times 10^5$ bit/s)&13.1 & 1.1\\
$P_u$ (mW)& 0.14 & 0.04\\
$P_d$ (mW)&0.16 &0.06 \\
\hline
\end{tabular}
\label{table:samle_statistics} 
\end{table}

\subsection{Parameter Estimation for GEV Distribution}
\label{inf}

Based on the statistics collected in Section \ref{expsetup}, we calculate the communication parameters, i.e., $z_{\epsilon_m}^u$, $z_{\epsilon_m}^d$, $\theta_u$ and $\theta_d$, which involve the inference of GEV distribution
of $\V$, $J_{(k)}$, and $H_{(k)}$. Since all of them follow the same process, we only show the process of inferring $\V$. We use tPacketCapture and WIFI SNR to capture a series of independent samples of real-time queue length and SNR, and then form a set $\{V_1,V_2,\cdots,V_N\}$, where $V_n = \frac{Q_n^u+o_n}{R_u}$, $o_n$ is a generated matrix data size measured in bits, $Q_n^u$ is the queue length when transmitting $o_n$ to the edge server and is determined based on the time tag when the matrix transmission command is invoked and executed, and $R_u$ is the corresponding bit rate. The samples are blocked into $\frac{N}{k}$ sequences of length $k$, for some large values of $k$, generating a series of block maxima, ${\V}_1,{\V}_2,\cdots,{\V}_{\frac{N}{k}}$, to which the GEV distribution for $\V$ in the case of transmitting data to the edge server is fitted. Please refer to \cite{coles2001introduction} for the details of using MLE to find the GEV distribution given series of block maxima. In the testbed experiment, we set $k$ (the order of statistics) as 1500, because by the end of executing the Smart Diagnosis, each device can generate about 1500 random matrices, transmit them to the server, and receive the   singular values. The data collection period is repeated for 100 times , which means $\frac{N}{k}=100$.

\begin{figure}[htb]
 \begin{center}
  \begin{tabular}{cc}
    \includegraphics[width= .47\columnwidth]{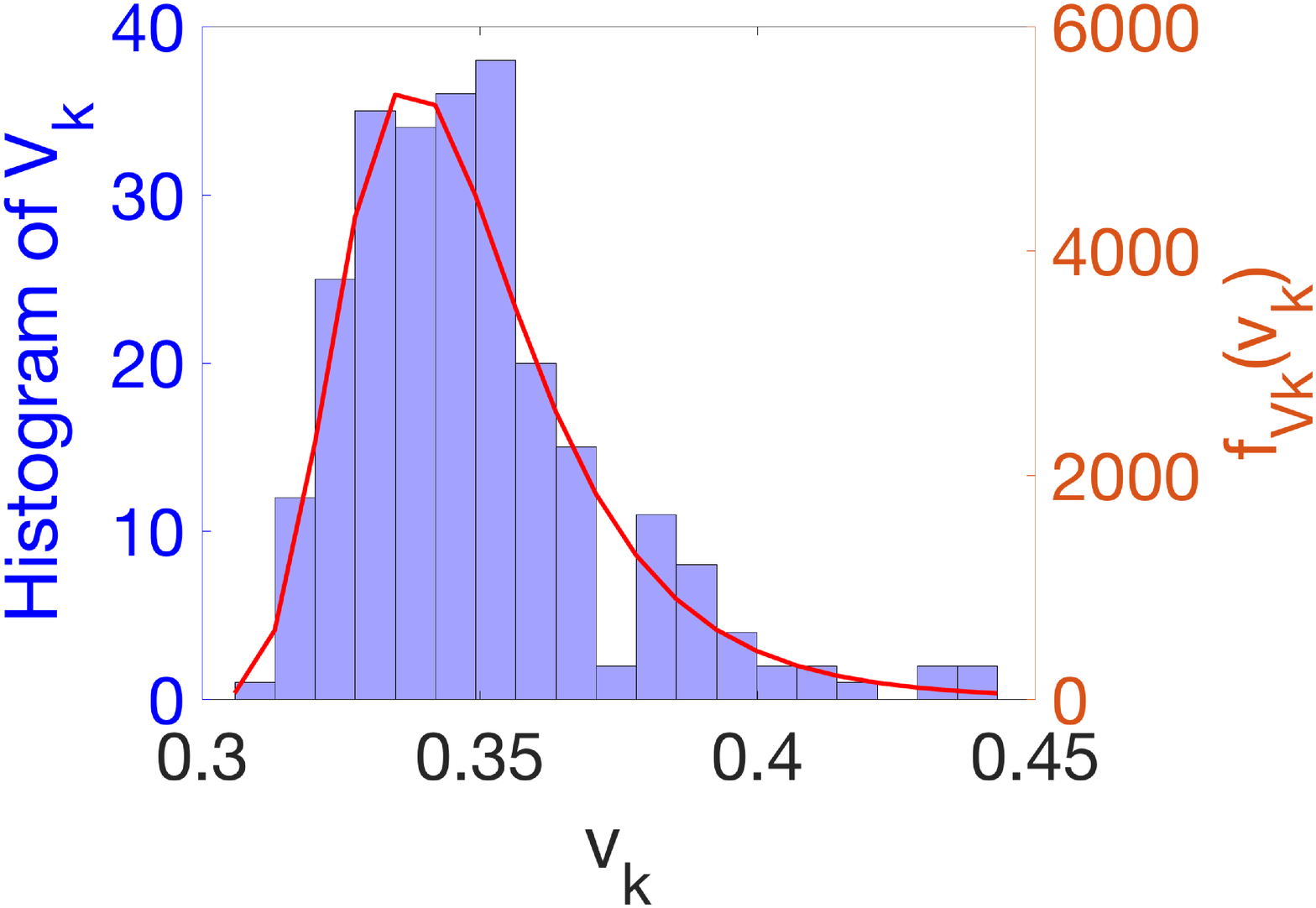}
		&
    \includegraphics[width= .47\columnwidth]{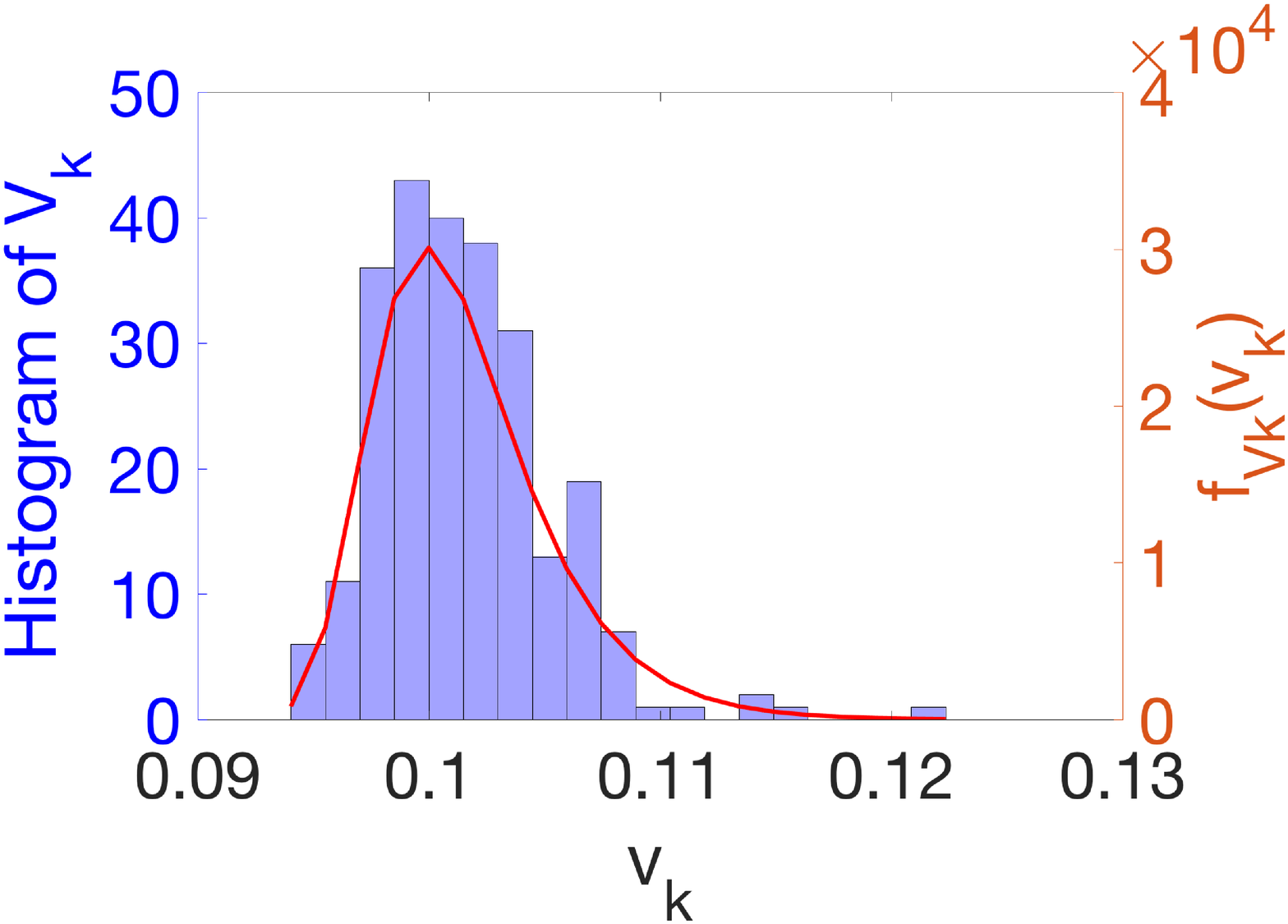}
    \\
    {\small  (a) GEV distribution of}  &
    {\small  (b) GEV distribution of}
        \\
    {\small  $V_{(k)}$ during data uploading}  &
    {\small  $V_{(k)}$ during data downloading}
    \\
        \includegraphics[width= .47\columnwidth]{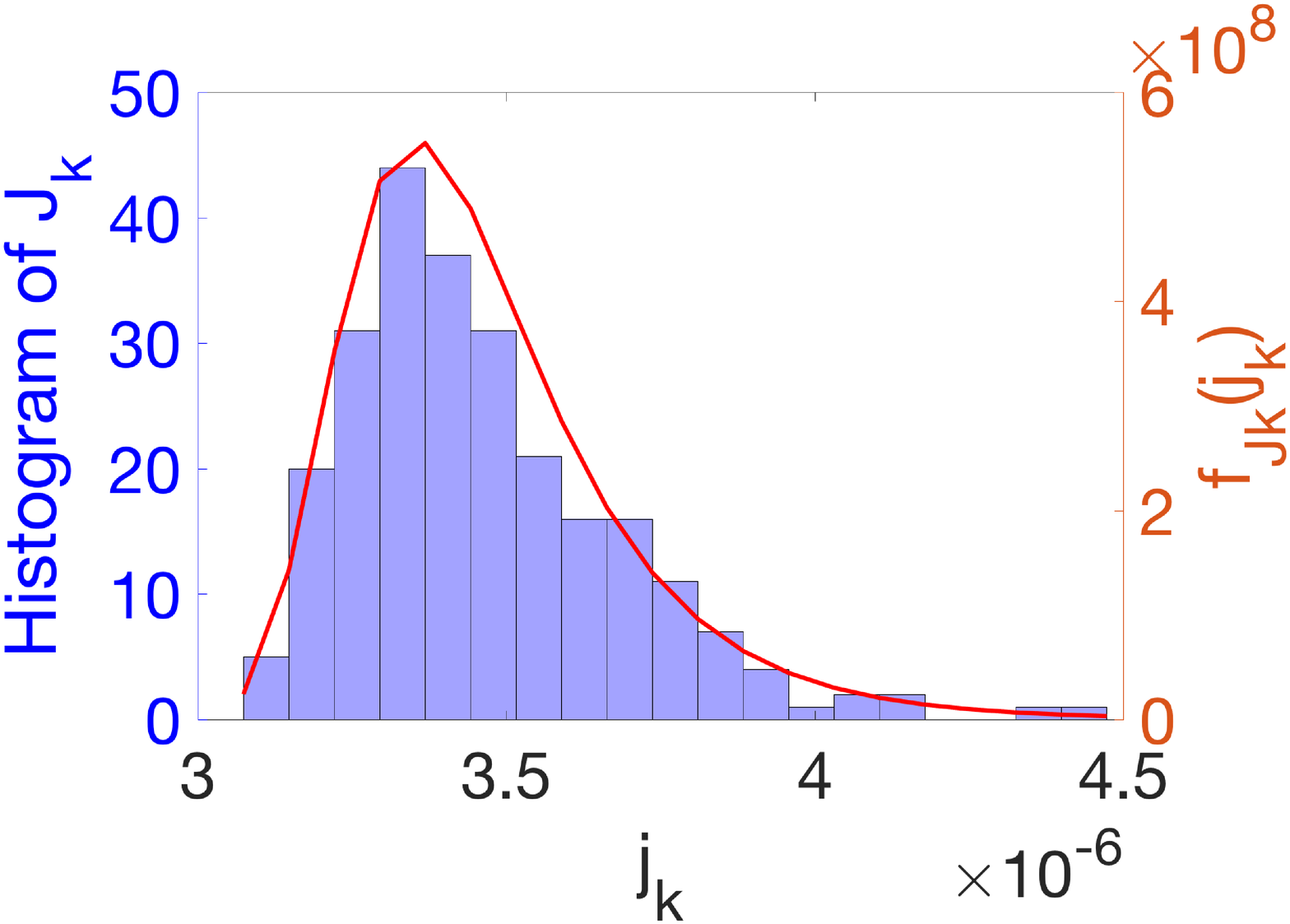}
		&
    \includegraphics[width= .47\columnwidth]{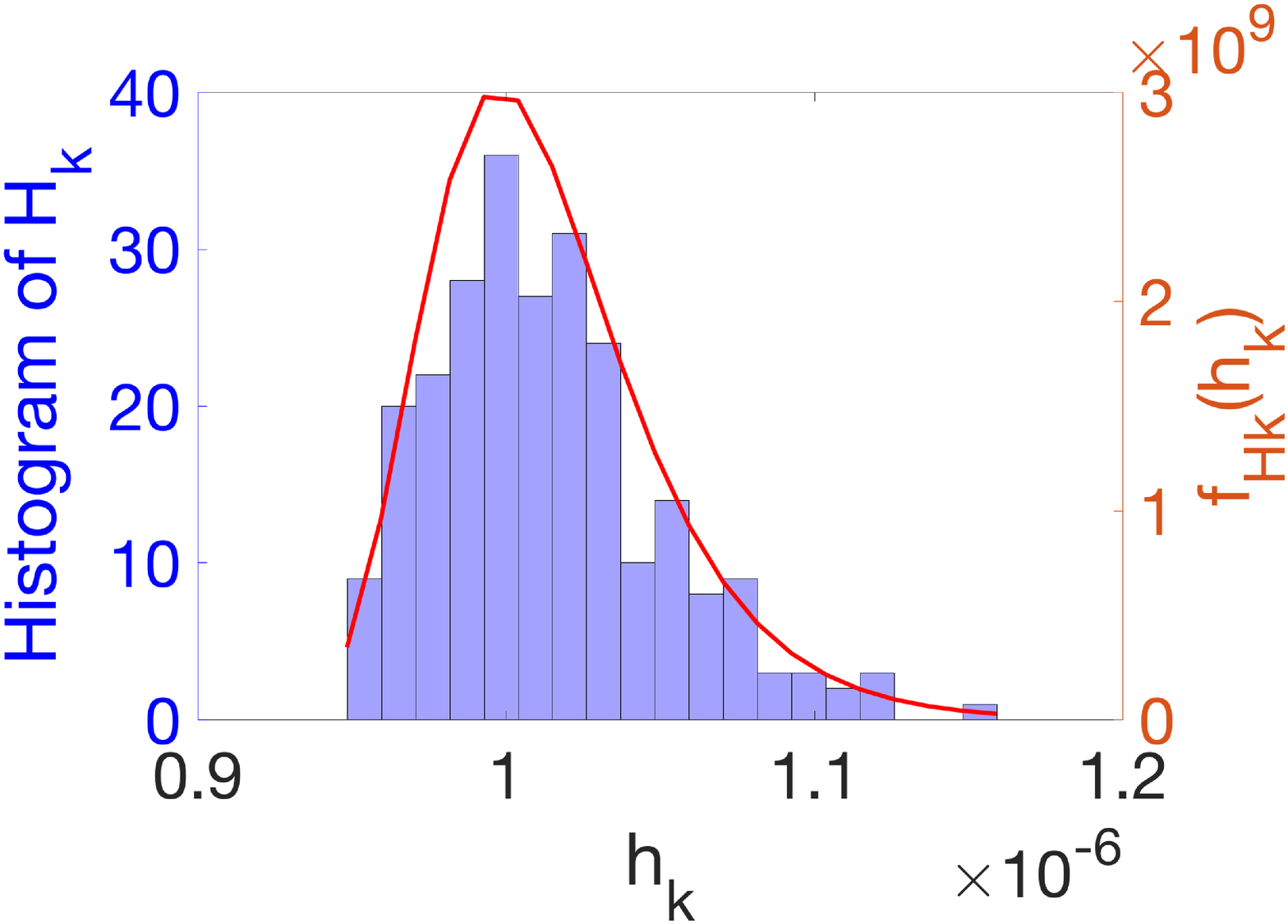}
    \\
    {\small  (c) GEV distribution of $J_{(k)}$}  &
    {\small  (d) GEV distribution of $H_{(k)}$}
        \end{tabular}
      \end{center}
 \caption{\label{fig:gev} Inferred GEV distributions.}
\end{figure}

Fig. \ref{fig:gev} shows the histogram of the block maximas (indicated by the blue bars) and the inferred GEV distributions (indicated by the red curves).
Specifically, by fixing $\epsilon_m^u = \epsilon_m^d = 0.1$, according to (\ref{z}), we calculate the $90\%$ quantile of $f_{V_k}(v_k)$ for transmitting data to the edge server and client device as $z_{\epsilon_m}^u=0.349$ and $z_{\epsilon_m}^d = 0.107$, respectively. By applying (\ref{mean}), we calculate $\theta_u = 4.81\times 10^{-4}$ and $\theta_d = 1.11\times 10^{-5}$. The parameters used for executing Smart Diagnosis in the MEC system are summarized in Table \ref{para}.


\begin{table*}[h]
\caption{The Summary of Experiment Settings.} 
\centering 
\begin{tabular}{c|c| c| c| c|c|c| c| c| c} 
\hline 
    $f_c$                   &  $f_s$              &$\kappa$& $z_{\epsilon_m}^u$ & $z_{\epsilon_m}^d$  & T&$\Delta$&$\theta_u$ & \multicolumn{2}{c}{$\theta_d$} \\ [0.5ex] 
    \hline
       $1.5$GHz & $2.4$GHz  &$10^{-24}$mW$/Hz^3$    & $0.349$s           &$0.107$s    & $0.5\times 10^4$ms&1&$4.81\times 10^{-4}$mW$\cdot$s/bit & \multicolumn{2}{c}{$1.11\times 10^{-5}$mW$\cdot$s/bit}    \\
\hline
\end{tabular}
\label{para} 
\end{table*}

\subsection{Performance of the Proposed Scheme}
\label{exp:performance}
To evaluate the efficacy of our proposed scheme, we calculate the offloading percentage and the  energy consumption   of Smart Diagnosis     using different computation offloading decisions obtained under  varying approximation parameter ($\epsilon$), while fixing the probabilities ($\epsilon_m^u$ and $\epsilon_m^d$) of the extreme case events happening during the period of data uploading and downloading. 

\begin{figure}[htb]
  \begin{center}
   \begin{tabular}{cc}
         \includegraphics[width= .45\columnwidth]{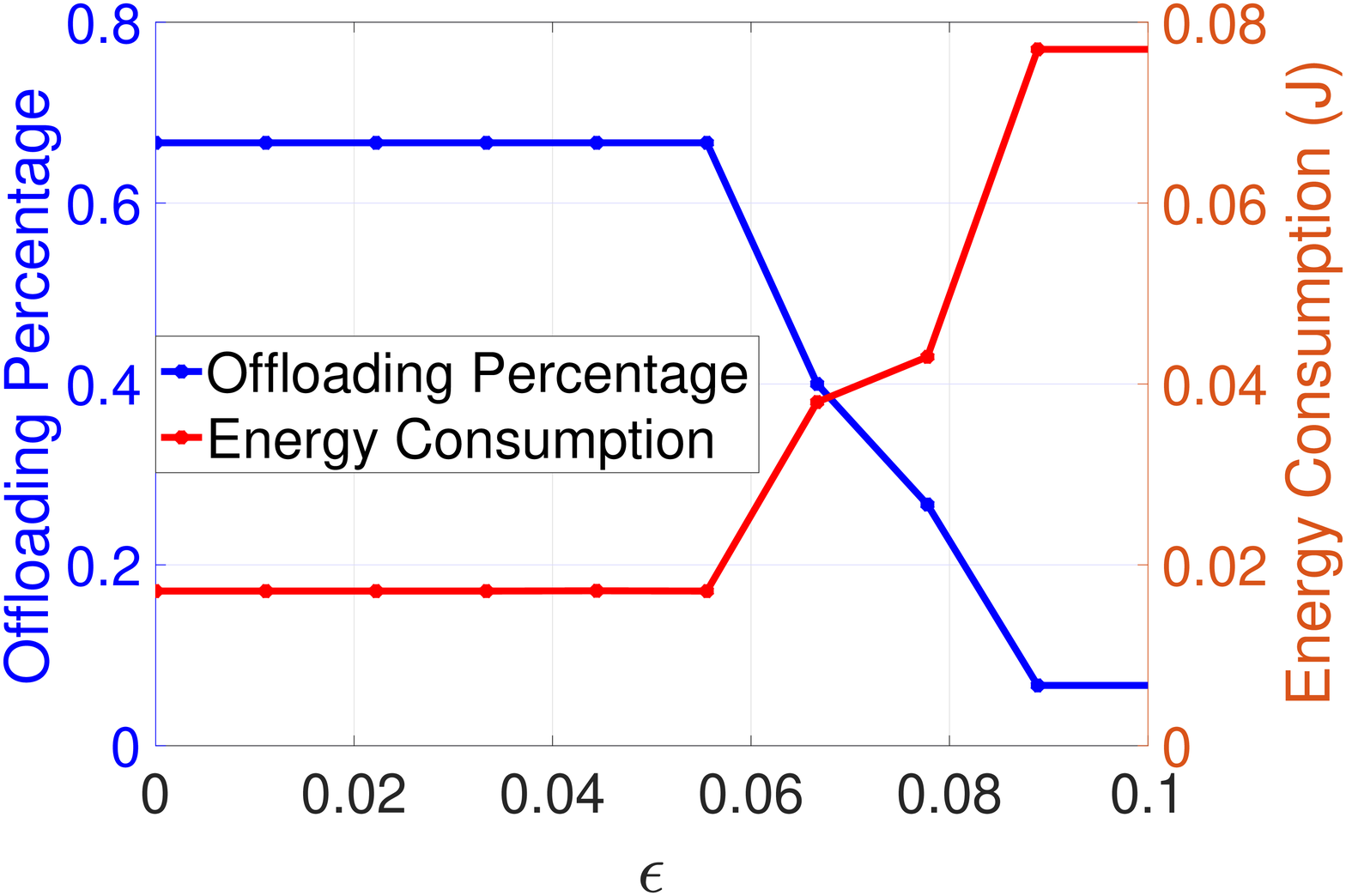}
		&
     \includegraphics[width= .45\columnwidth]{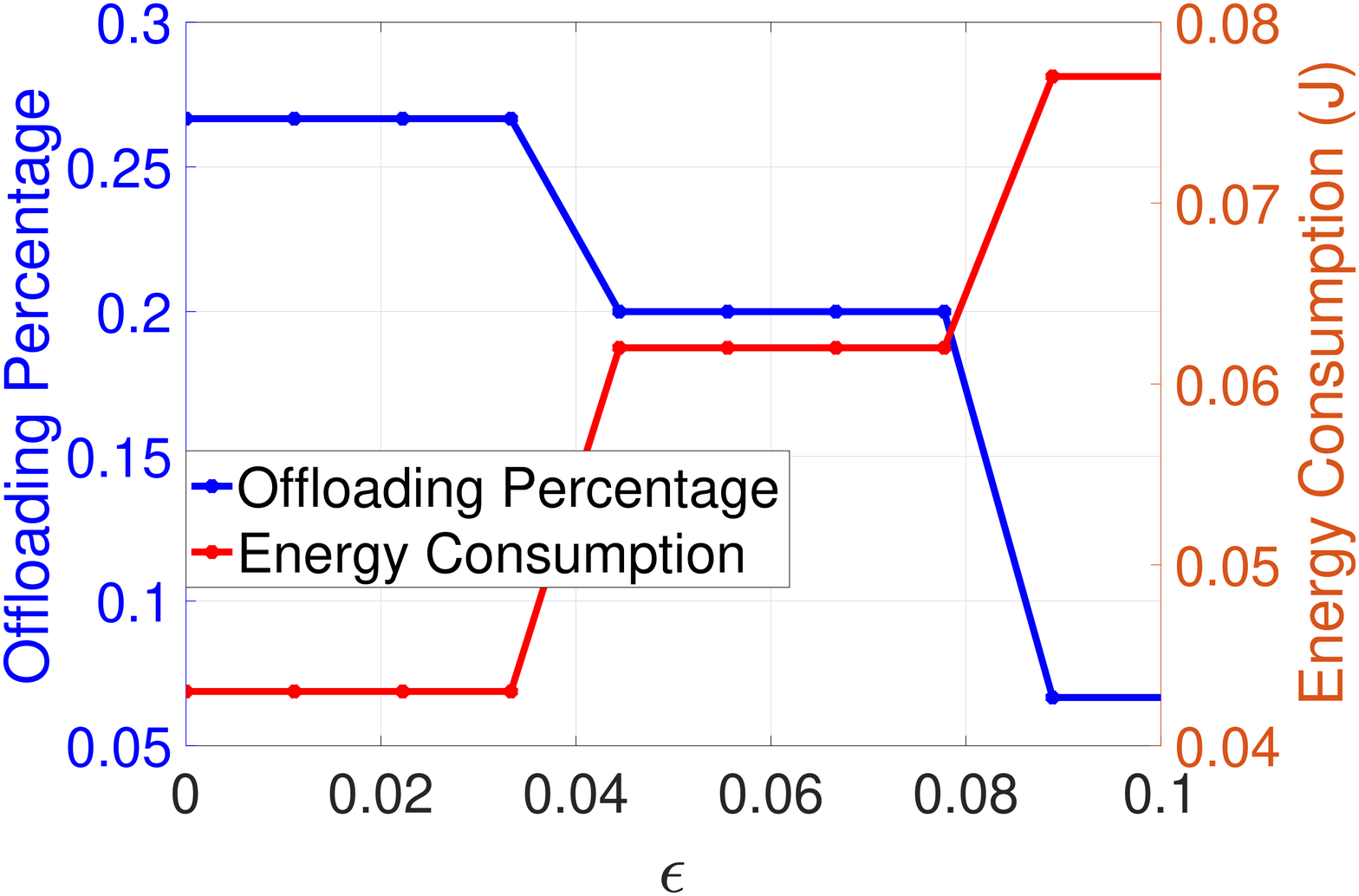}
     \\
         {\small  (a) Fix $\epsilon_m^d = \epsilon_m^u = 0.01$.}  &
    {\small  (b) Fix $\epsilon_m^d = \epsilon_m^u = 0.1$.}
    \\
              \includegraphics[width= .45\columnwidth]{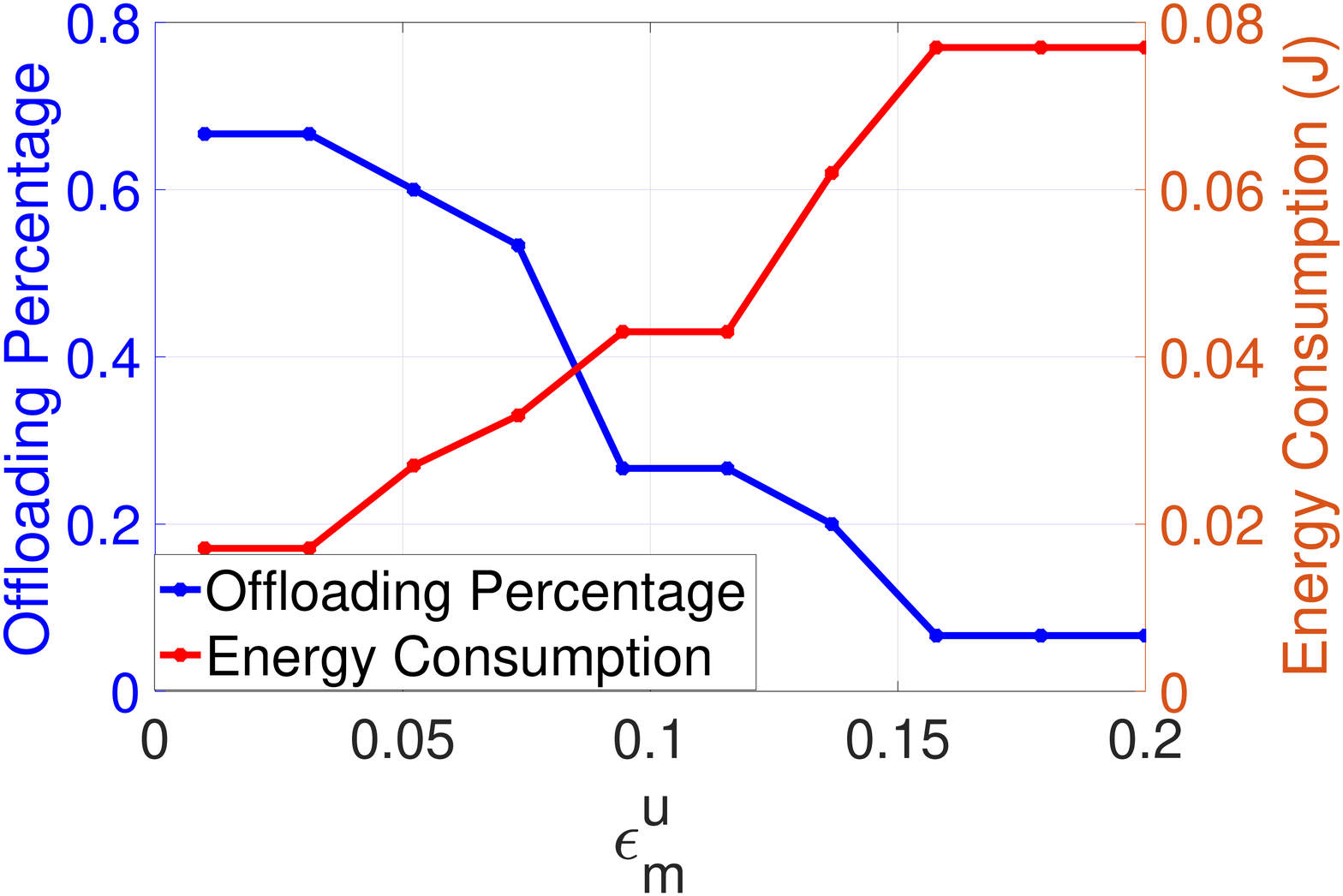}
		&
     \includegraphics[width= .45\columnwidth]{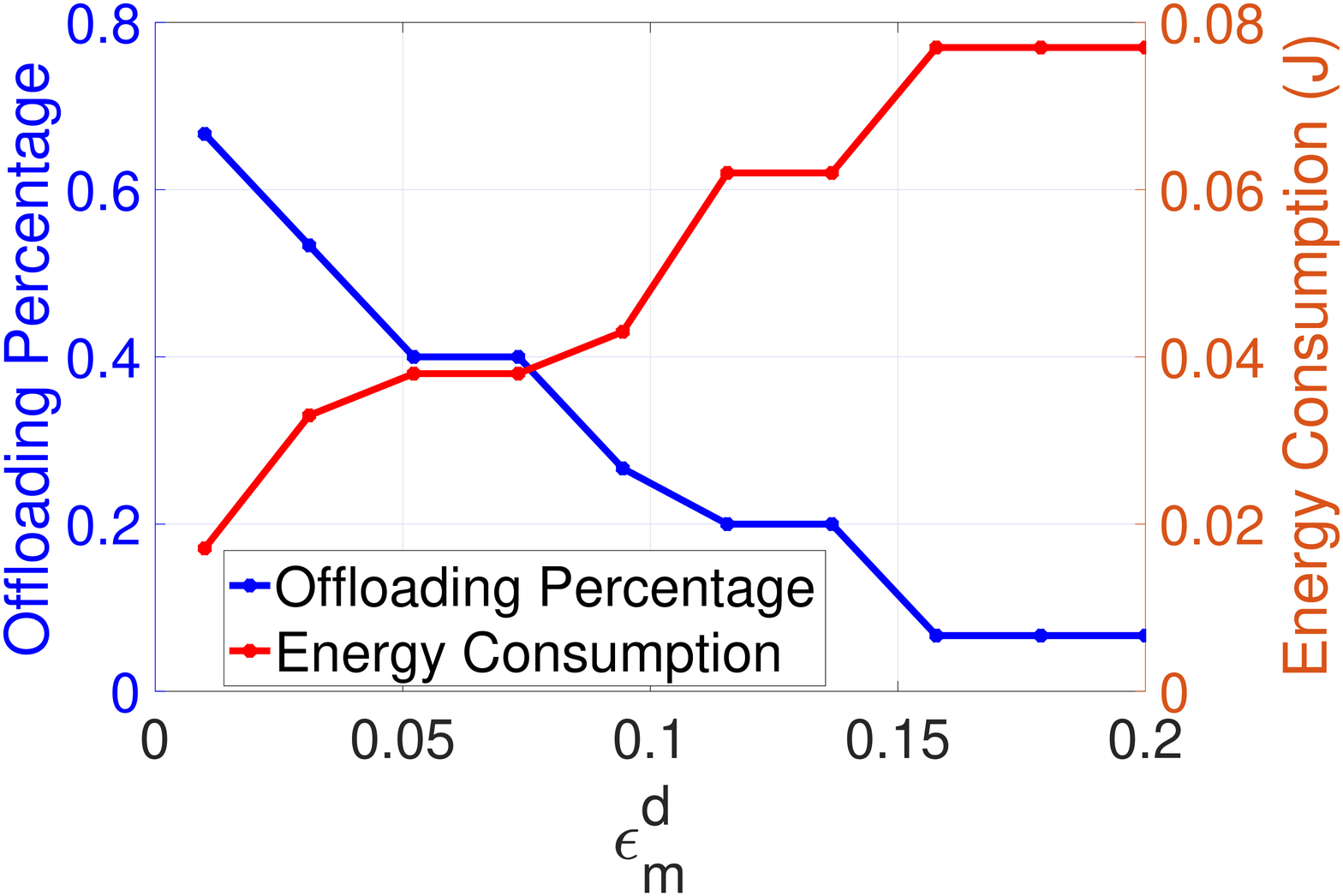}
     \\
    {\small  (c) Fix $\epsilon = 0.03, \epsilon_m^d = 0.01$.}  &
    {\small  (d) Fix $\epsilon = 0.03, \epsilon_m^u = 0.01$.}
        \end{tabular}
       \end{center}
  \caption{\label{fig:approx} Offloading percentage and energy consumption under various approximation parameter and queueing delay.} 
\end{figure}

We show the   results in Fig. {\ref{fig:approx}}, where the blue dotted line represents the offloading percentage and the red dotted line is the energy consumption. Specifically, Fig. {\ref{fig:approx}}(a) and (b) are the results of setting $\epsilon_m^d = \epsilon_m^u = 0.01$ and $\epsilon_m^d = \epsilon_m^u = 0.1$, respectively.\footnote{In practice, we can choose large values of $\epsilon_m^d$ and  $\epsilon_m^u$ if there are multiple mobile devices utilizing the computation resource at the edge, because in this case, the probability of experiencing high queuing size and low communication   rate will be higher.} We  see that our scheme is very robust under various $\epsilon$. For example, in Fig. {\ref{fig:approx}}(a), when $\epsilon_m^d = \epsilon_m^u = 0.01$, the offloading policy is optimal as long as the approximation parameter does not exceed $0.05$. This is because the client device favors an energy-friendly computation offloading policy that offloads more computing nodes, and when the extreme case events happen with a lower probability, the radio channel is able to support more data transferring between the client device and the edge server. Even increasing the probability that extreme case events occur, we can still obtain an optimal offloading policy when $0\leq \epsilon \leq 0.03$ as shown in Fig. {\ref{fig:approx}}(b). Based on the obtained optimal offloading decision for Smart Diagnosis when $\epsilon = 0.03, \epsilon_m^d = \epsilon_m^u = 0.1$, modules named ``SURF Description'', ``SIFT Description'', ``KNN1'', ``KNN2", ``Tokenization'', and ``POS Tagging'' in Fig. \ref{fig:tag} are offloaded.

We further evaluate the impact of queuing delay on the performance by varying probabilities $\epsilon_m^u$ and $\epsilon_m^d$, and show the results in Fig. \ref{fig:approx} (c) and (d). We can see that for both cases, the offloading percentage decreases and $\Psi$ increases with the increasing values of $\epsilon_m^u$ and $\epsilon_m^d$. The reason is that as $\epsilon_m^u$ and $\epsilon_m^d$ increase, the channel conditions and queuing delay cause the increase in data transmission time, thus the proposed scheme decreases the number of modules to be offloaded.

\subsection{Performance Degradation Caused by Simple Network Condition and Queueing   Assumptions}
\label{exp:performance_degrade}

Having demonstrated the efficacy of the proposed scheme, we now empirically corroborate that it will lead to higher  energy consumption for the client device if we simply adopt   strong assumptions on the network conditions or the  queuing delays when deciding which task module to offload. Specifically, we consider the offloading decisions for the Smart Diagnosis   obtained under the 
 following   scenarios: (i) a block-fading communication channel, and (ii)  constant queuing size of the out-going  buffers at the  client device and server.

\noindent\textbf{Scenario (i).} 
The block-fading channel assumes that the channel condition does not change over the duration of application execution, thus, according to the  Shannon-Hartley theorem, the client device can have  constant uploading and downloading bit rates (i.e., $R_u$ and $R_d$). As a result, the  unknown conditions when making computation offloading decisions are $Q_{m,n}^u$, $Q_{m,n}^d$ (cf. (\ref{dependence})), $P_u$, and $P_d$ (cf. (\ref{energu})). The offloading decision can still be obtained by the developed column generation based algorithm by considering constant $R_u$ and $R_d$ when fitting the  required GEV distributions in Section \ref{inf}.

%

\begin{figure}[!htb]
    \centering
  \begin{center}
   \begin{tabular}{cc}
         \includegraphics[width= .45\columnwidth]{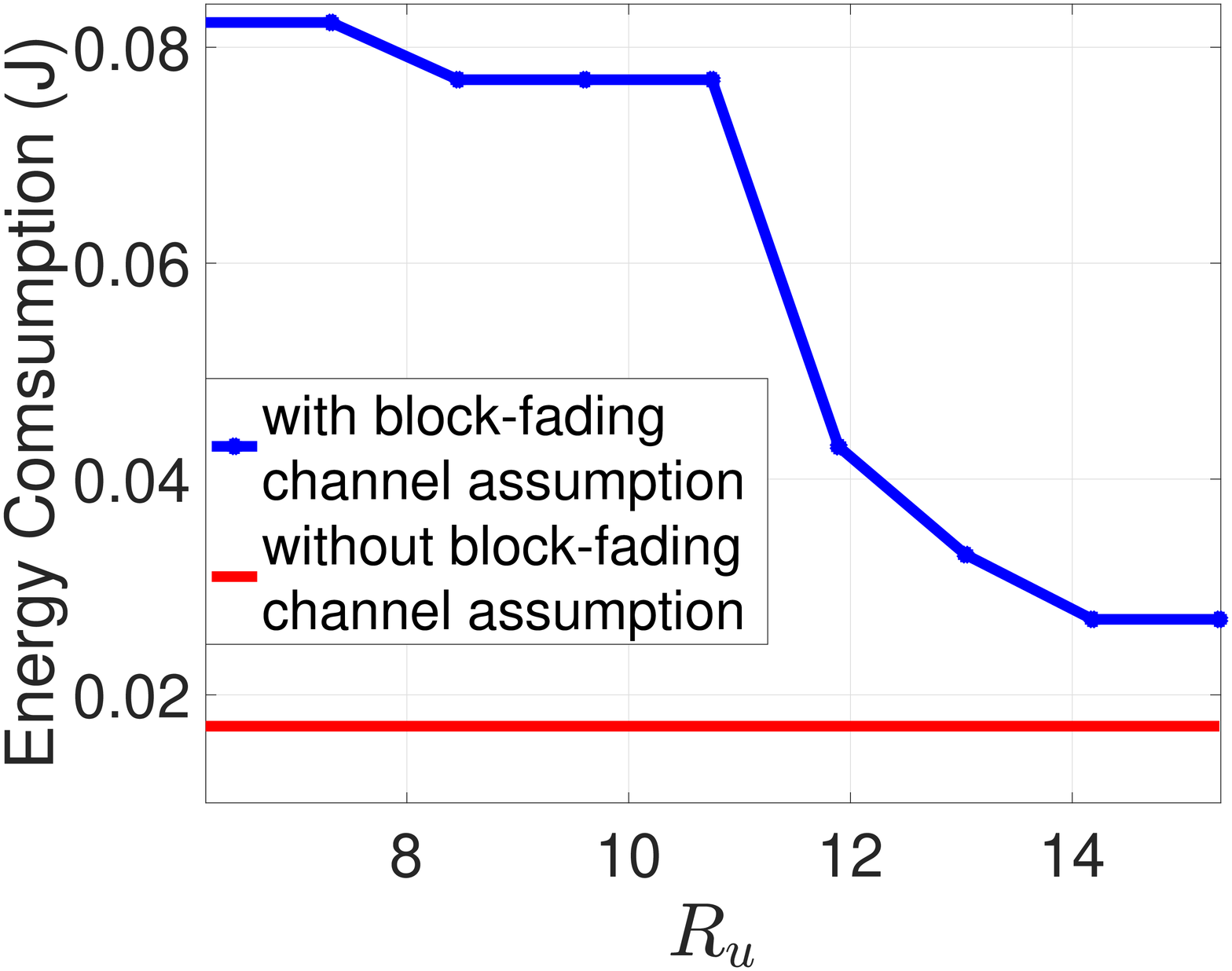}
		&
     \includegraphics[width= .45\columnwidth]{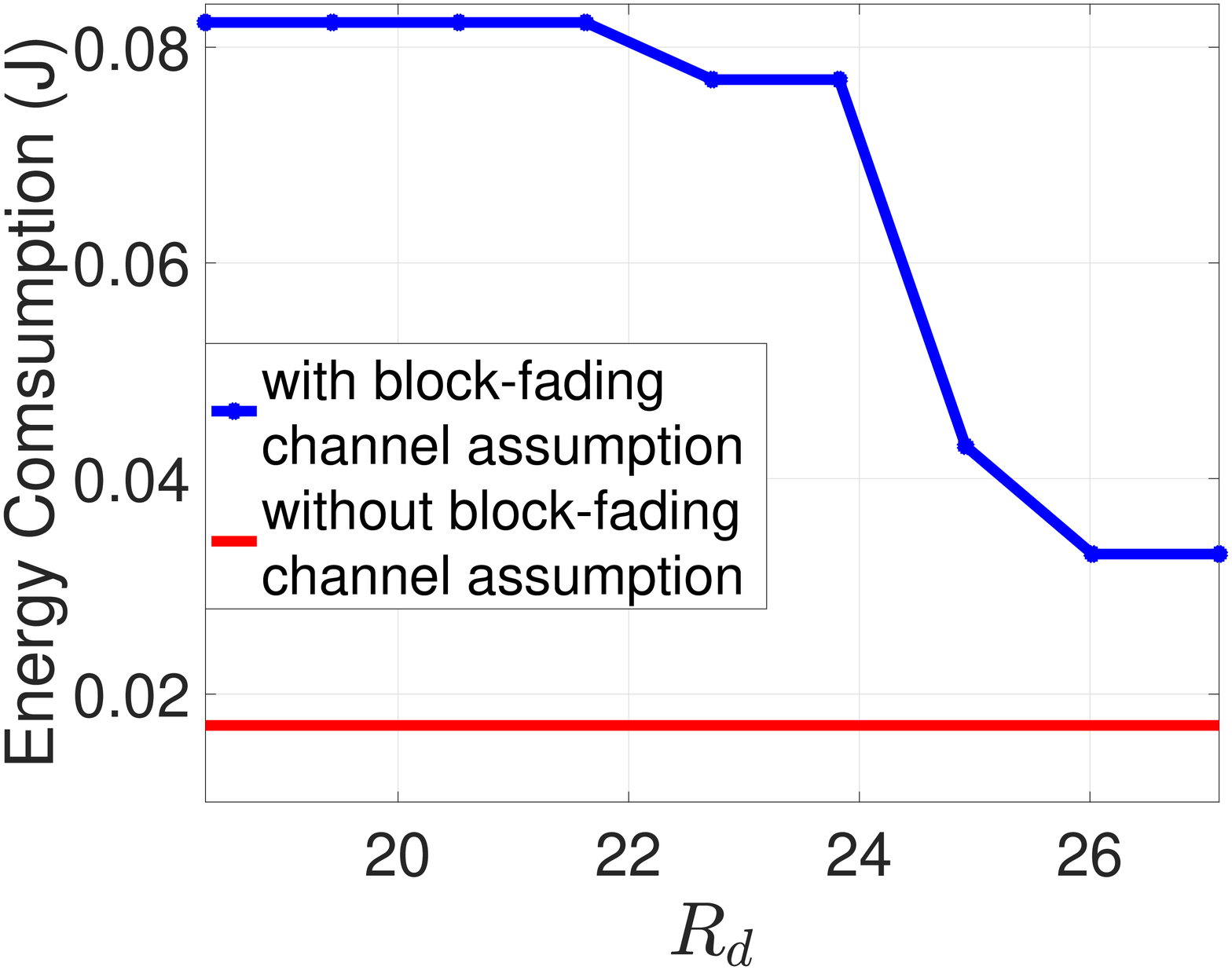}
     \\
    {\small  (a) $R_d = 22.7$MB/s, vary $R_u$}  &
    {\small  (b) $R_u = 10.7$MB/s, vary $R_d$}
        \end{tabular}
       \end{center}
  \caption{\label{fig:scenario_i} Energy consumption comparison between offloading decisions obtained with and without the block-fading channel assumption.} 
    \end{figure}

In Fig. ~\ref{fig:scenario_i} (a) and (b), we present the empirical energy consumption of the client device when executing Smart Diagnosis using the offloading decisions obtained with  the block-fading channel assumption (the blue curves).  We also plot the energy consumption obtained  without assuming block-fading channel (the red lines) for comparison.  In this experiment, we fix  $\epsilon_m^d = \epsilon_m^u = 0.01$ and $\epsilon=0.05$. More specifically, in Fig. ~\ref{fig:scenario_i} (a), we fix $R_d$ to be the average value computed using SNR measured by WiFiSNR in Section \ref{expsetup}, and vary $R_u$ in the range of  $\mathbb{E}[R_u]\pm4.5$MB/s, where $\mathbb{E}[R_u] = 10.7$MB/s is   obtained from Section \ref{expsetup}. 
Similarly, in Fig. ~\ref{fig:scenario_i} (b), we fix $R_u$ to be the average value obtained  in Section \ref{expsetup}, and vary $R_d$ in the range of  $\mathbb{E}[R_d]\pm4.5$MB/s, where $\mathbb{E}[R_d] = 22.7$MB/s is  also obtained from Section \ref{expsetup}.  Clearly, by considering the uncertainty in downloading and uploading bit rates, i.e., without the strong assumptions on the communication channel, the offloading decisions achieved by our proposed scheme can always lead to much lower empirical energy consumption.  The reason is that under the block-fading channel assumption, when $R_u$ or $R_d$ is low,   the offloading decision will assign more task module at the client device, which increases its energy consumption. On the other hand, when $R_u$ or $R_d$ is high,  the offloading decision will assign more task module to  the server. However, in practice, it may take the client device more time to transmit or receive the data if the data rates drop in the course of  the application execution, because we carry the mobile device and walk at  various speed during all the experiments. Similar phenomenon  has also been shown in    \cite{lapidoth1998reliable}, i.e.,  the varying locations of the local mobile transmitter and receiver can lead to uncertainties in channel law as well as arbitrarily varying communication bit rates.

\noindent\textbf{Scenario (ii).}  In this case, the  unknown conditions in our  computation offloading scheme are $R_u$, $R_d$, $P_u$, and $P_d$. The offloading decision is obtained using the   column generation based algorithm by considering the constant $Q^u_{m,n}$ and $Q^d_{m,n}$.  

    \begin{figure}[!htb]
  \begin{center}
   \begin{tabular}{cc}
         \includegraphics[width= .45\columnwidth]{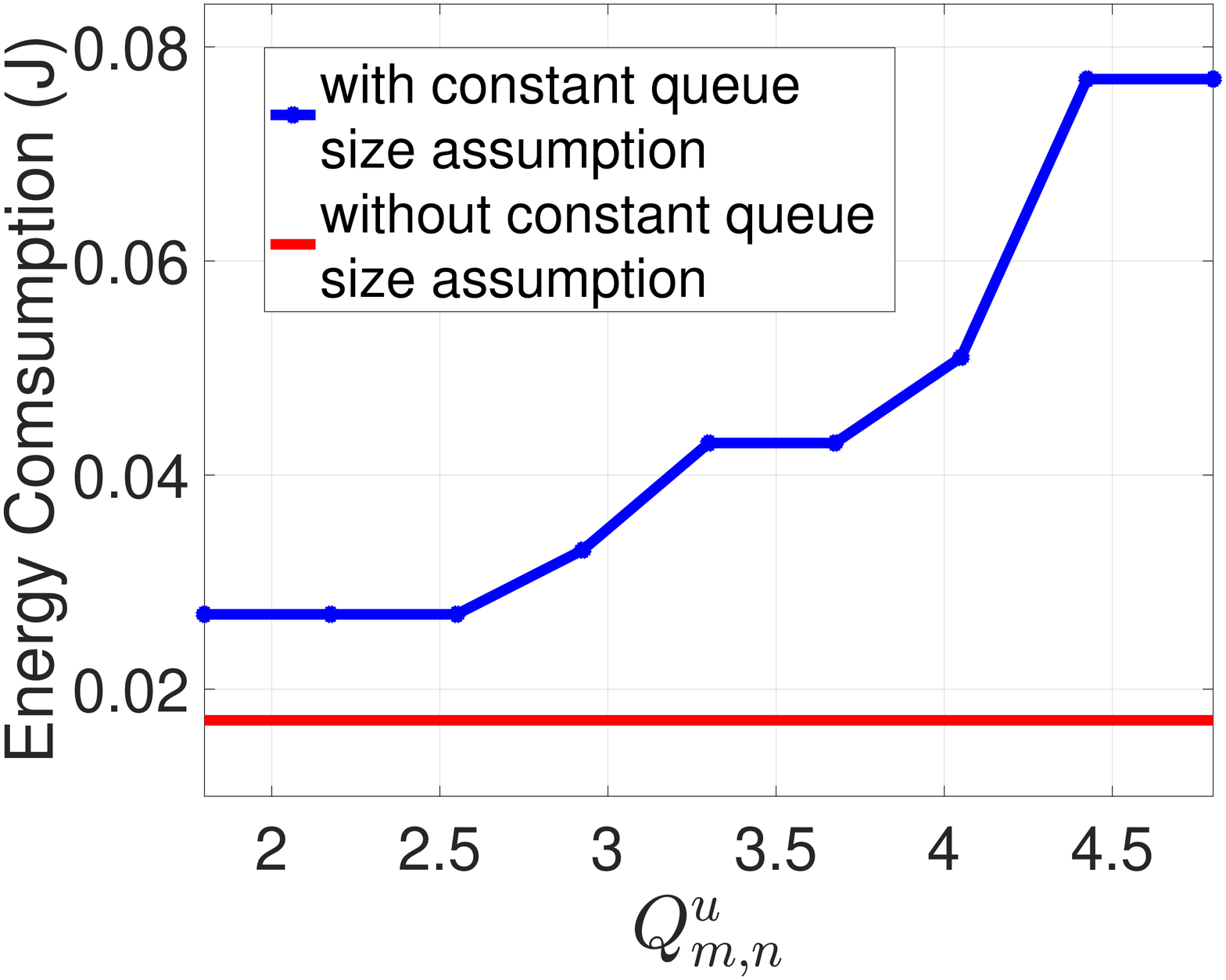}
		&
     \includegraphics[width= .45\columnwidth]{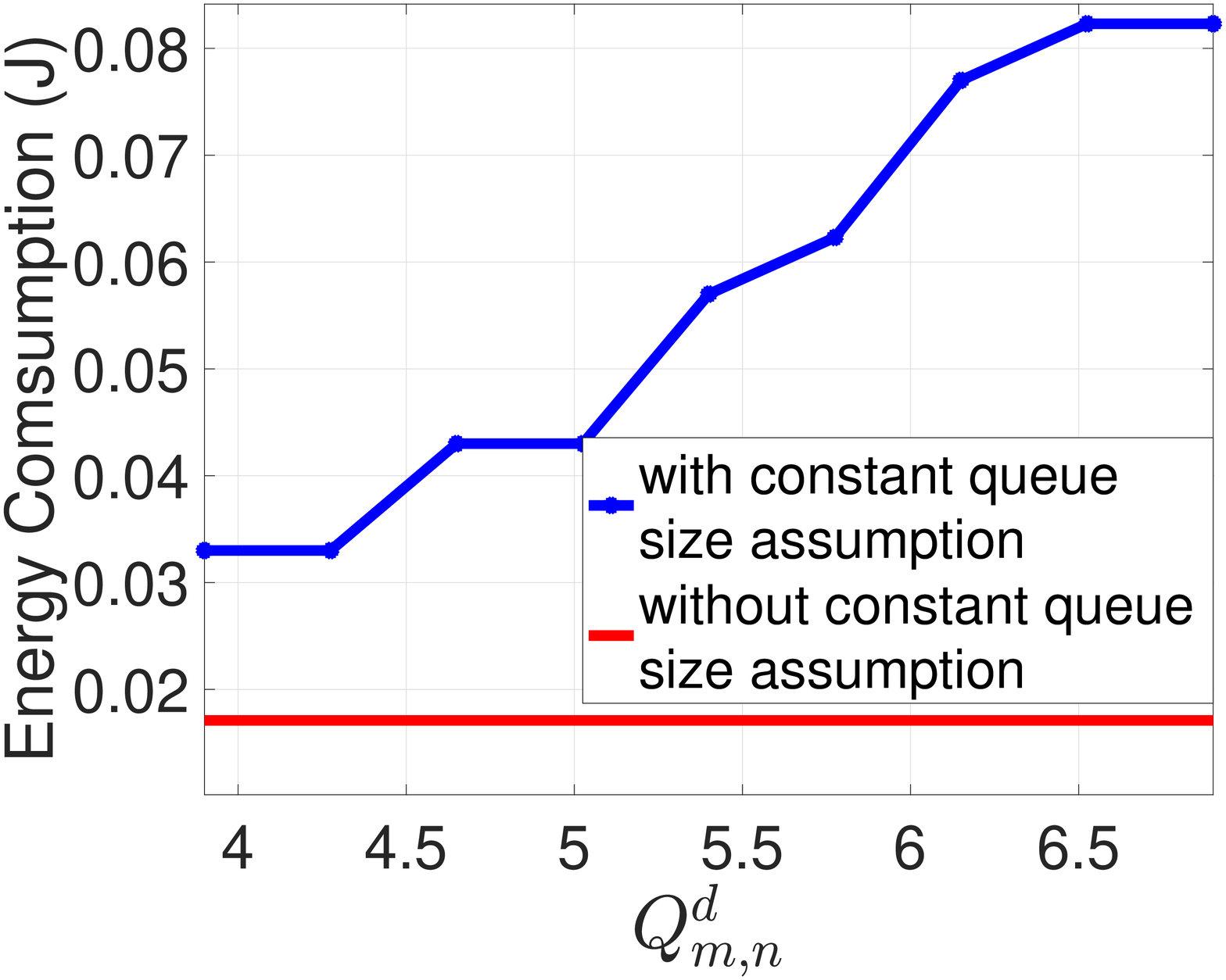}
     \\
    {\small  (a) $Q^d_{m,n} = 3.3$KB, vary $Q^u_{m,n}$}  &
    {\small  (b) $Q^u_{m,n} = 5.4$KB, vary $Q^d_{m,n}$}
        \end{tabular}
       \end{center}
  \caption{\label{fig:scenario_ii} Energy consumption comparison between offloading decisions obtained with and without the constant queuing size assumption.} 
\end{figure}

In Fig. ~\ref{fig:scenario_ii} (a) and (b), we present the empirical energy consumption of the client device when executing Smart Diagnosis using the offloading decisions obtained with  the constant queuing size assumption (the blue curves).  Additionally, the energy consumption  (the red lines) obtained  without such assumption is shown  for comparison.  In this experiment, we still fix  $\epsilon_m^d = \epsilon_m^u = 0.01$ and $\epsilon=0.05$. Particularly, in Fig. ~\ref{fig:scenario_ii} (a), we fix $Q^d_{m,n}$ to be the average value  measured by tPacketCapture in Section \ref{expsetup}, and vary $Q^u_{m,n}$ in the range of  $\mathbb{E}[Q^u_{m,n}]\pm1.5$KB, where $\mathbb{E}[Q^u_{m,n}] = 3.3$KB is   obtained from Section \ref{expsetup}. 
Similarly, in Fig. ~\ref{fig:scenario_ii} (b), we fix $Q^u_{m,n}$ to be the average value obtained  in Section \ref{expsetup}, and vary $Q^d_{m,n}$ in the range of  $\mathbb{E}[Q^d_{m,n}]\pm1.5$MB/s, where $\mathbb{E}[Q^d_{m,n}] = 5.4$MB/s is  also obtained from Section \ref{expsetup}.  Clearly, by considering the uncertainty of the queuing size, i.e., without the constant queuing size assumption, the offloading decisions achieved by the proposed scheme can always lead to much lower empirical energy consumption.  Similar analysis also applies to this scenario; under the assumption of constant queuing size at the out-going buffer of the client device and the server, when $Q^u_{m,n}$ or $Q^d_{m,n}$ is low,   the offloading decision will assign more task modules to the server to save local energy consumption, but it may take the client device more time to transmit or receive the data, which incur extra communication energy, if the queuing size    increase during the application execution. The reason is that  sudden burst of serving requests can overflow the edge servers' resource and may even cause MEC failure \cite{satria2017recovery,babou2018home}. On the other hand,  if $Q^u_{m,n}$ or $Q^d_{m,n}$ is high,  the offloading decision will assign more task modules to  the local device and increase its energy consumption.


\subsection{Performance Improvement  over Existing     Schemes}
\label{exp:compare}

We compare our proposed algorithm with two state-of-the-art offloading approaches, i.e., Hermes \cite{kao2017hermes} and JSCO \cite{mahmoodi2016optimal} discussed in Section \ref{RW}. 
The experiment   are conducted by varying $\epsilon$ when $\epsilon_m^d = \epsilon_m^u = 0.01$, and the offloading percentage and energy consumption are calculated for all algorithms. The   results are shown in Fig. \ref{fig:compare_soa}. 
We   see that our scheme can always lead to the lowest energy consumption by offloading more computation modules as long as the approximation parameter $\epsilon$ is less than $0.05$. Particularly, when $\epsilon\leq 0.05$, our     scheme can save more than $50\%$ energy for the local device, as the energy consumption of executing Smart Diagnosis using Hermes, JSCO, and our scheme are about $0.03J$, $0.028J$, and $0.014J$, respectively. The reason is that JSCO fails to consider   uncertainties in dynamic radio channels with queueing delay, and Hermes introduces extra energy consumption due to   communication overhead caused by continuously probing the channel.

\begin{figure}[h]
    \centering
    \includegraphics[width=.45\textwidth]{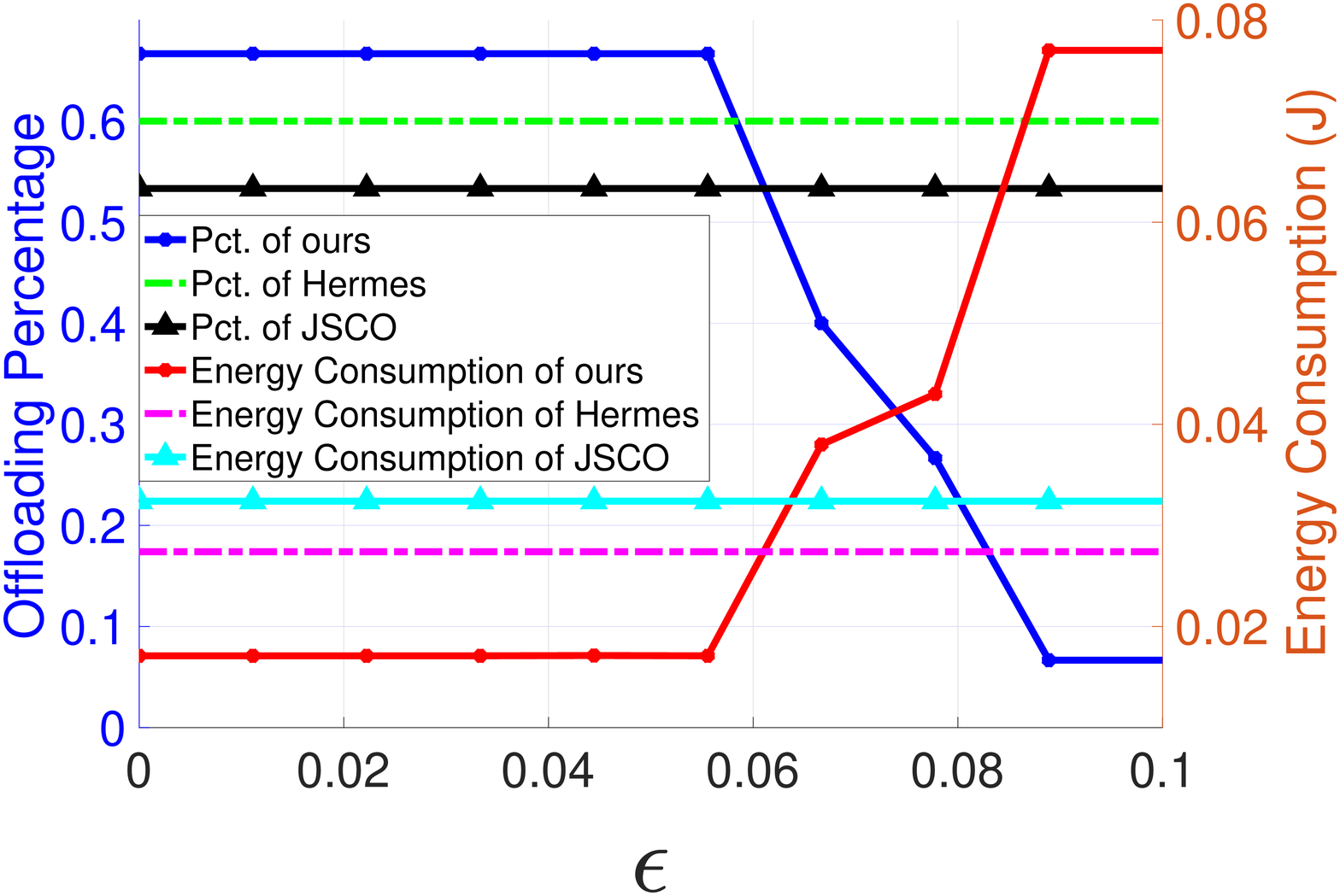}
    \caption{Performance comparison with state-of-the-art offloading approaches.} 
    \label{fig:cmp_soa}
\end{figure}


\subsection{The Scalability of The Proposed Scheme}\label{exp:scale}

To explore the scalability of the proposed scheme, we measure the number of offloaded nodes when running simulated DAGs with different sizes. Specifically, we simulate DAGs with layer-by-layer structure, where a random number of nodes is generated for each layer and edges are added between nodes across layers with a given probability $p$. In this experiment, the computation workload and output data size of each module in simulated DAGs are all attributed to half-normal distributions (nonnegative support) with different parameters. Besides, the communication parameters used in this simulation experiment  are the same as those in Table \ref{para}.

\begin{figure}[h]
    \centering
    \includegraphics[width=.45\textwidth]{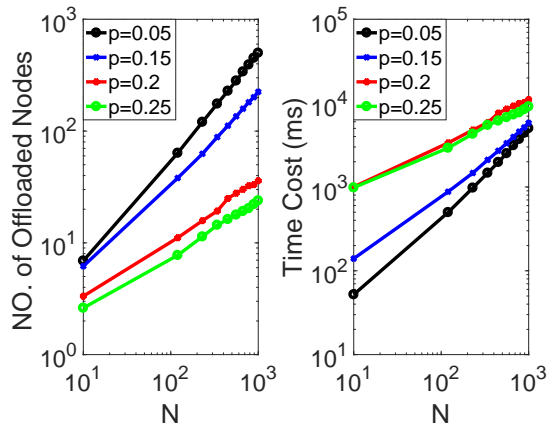}
    \caption{Evaluation of the scalability of the proposed computation offloading algorithm.} 
    \label{fig:compare_scale}
\end{figure}

We vary the number of nodes in a DAG from $10^1$ to $10^3$ and consider four connecting probabilities, i.e., $p\in\{0.05, 0.15, 0.2, 0.25\}$. In the left panel of Fig. \ref{fig:compare_scale}, we plot the number of offloaded nodes versus the size of DAG when having $\epsilon = 0.03$ and $\epsilon_m^u = \epsilon_m^d = 0.01$. We   see that by fixing a DAG size, i.e., $N=10^2$, the number of offloaded modules decreases with the increasing $p$. This is because the denser the DAG, the more dependency among modules, and offloading will cost more energy in dense DAGs than that in sparse ones. 
Moreover, we   also observe that by fixing a connecting probability, i.e., $p = 0.05$, the number of offloaded nodes grows linearly with the size of DAG. Since the proposed scheme identifies one column at each iteration, it scales linearly with the graph size. This indicates that our scheme is computationally efficient. In the right panel of Fig. \ref{fig:compare_scale}, we also plot the time cost 
of running our proposed algorithm on these simulated DAGs. We can see that the computation time also scales linearly with the graph size, and it suggests that our proposed algorithm can work in real-world MEC system. 


\subsection{Offloading Applications with Special  Dependency}
\label{exp:special}
In this section, to corroborate the findings in Theorem \ref{seq_dep} and \ref{para_dep}, we also simulate special applications with only sequential or parallel dependency. 
Specifically, we consider  an application composed of $14$ task modules. The computation workload and output data size for each task module are sampled from Poisson distributions with $\lambda_1 = 10^6$Hz and $\lambda_2 = 12$kb.

\begin{figure}[htb]
 \begin{center}
  \begin{tabular}{cc}
    \includegraphics[width= .45\columnwidth]{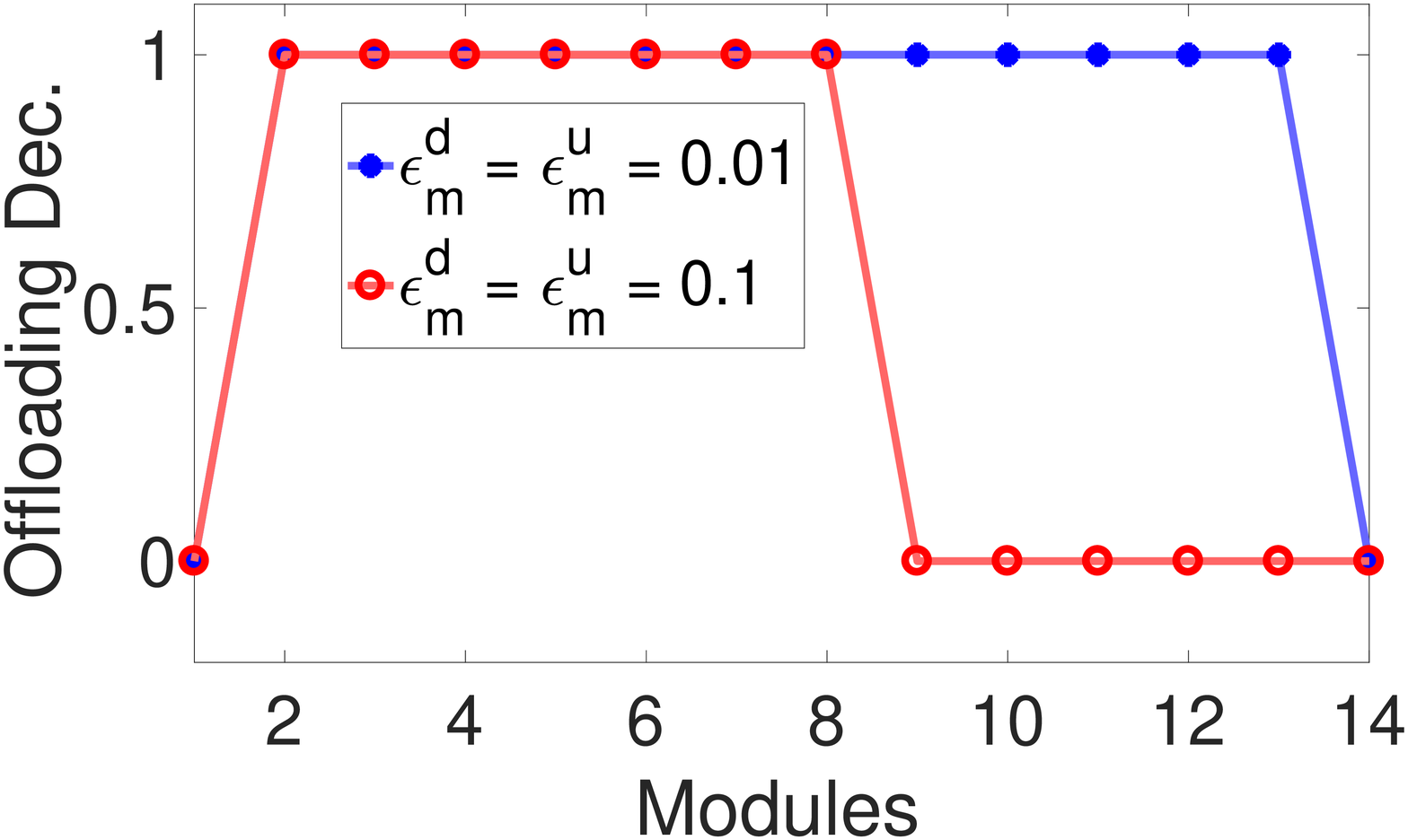}&
         \includegraphics[width= .45\columnwidth]{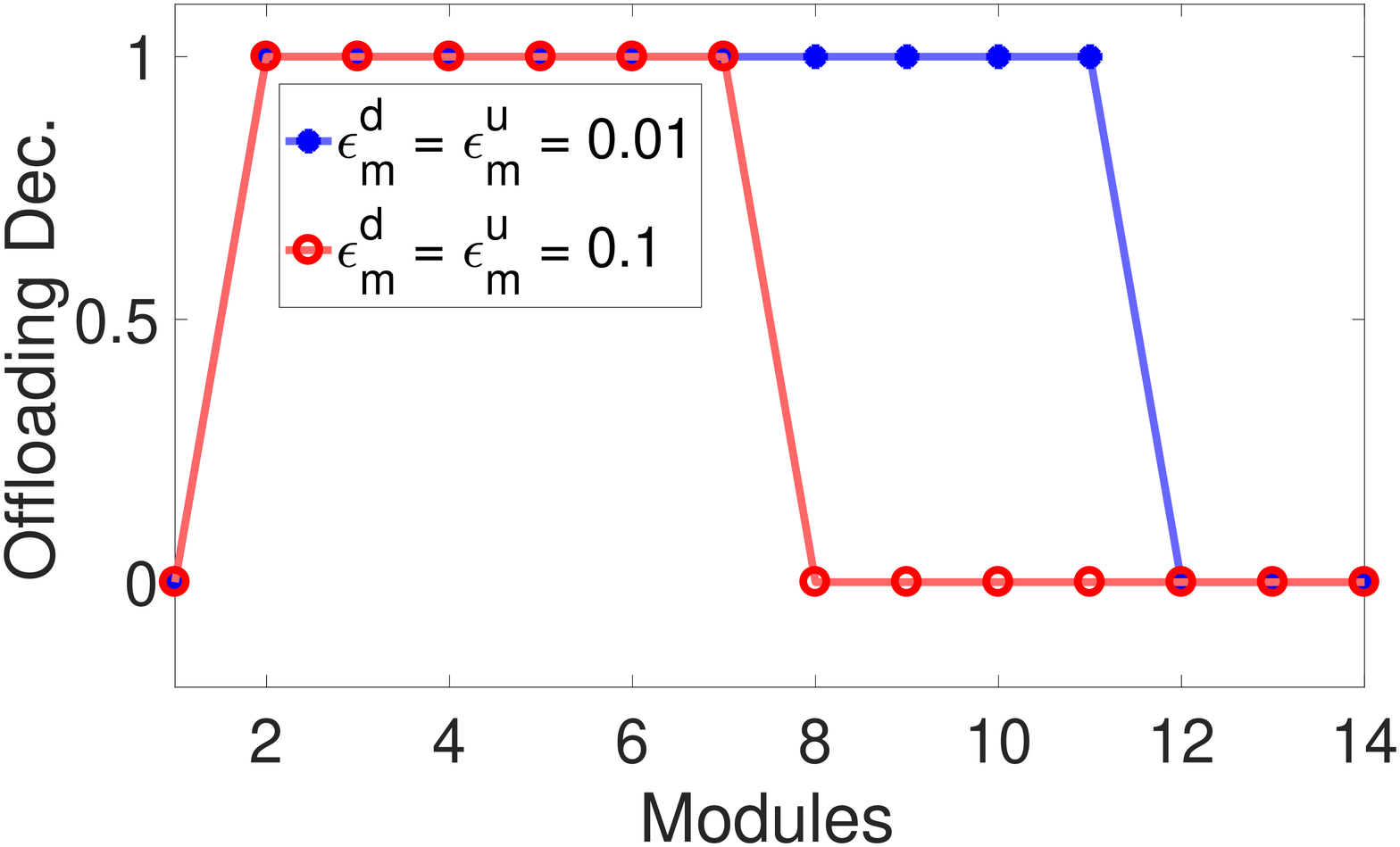}
	\\
	{\small  (a) $\epsilon=0$}
&
    {\small  (b) $\epsilon=0.03$}
        \end{tabular}
      \end{center}
 \caption{\label{fig:seq_e0} Offloading decision for sequential dependency DAG.}
\end{figure}

Fig. \ref{fig:seq_e0} shows the offloading decision of task modules with only sequential dependency. In particular, we compare the offloading decisions by considering two approximation parameters, i.e., $\epsilon = 0$ and $\epsilon = 0.03$, and for each $\epsilon$, we consider the effect of $\epsilon_m^d$ and $\epsilon_m^u$ on the offloading policy. In Fig. \ref{fig:seq_e0}, we use the blue dotted line to show the offloading policy under $\epsilon_m^d = \epsilon_m^u=0.01$, use the red circled line to show the offloading policy under $\epsilon_m^d = \epsilon_m^u=0.1$, and use $1$ or $0$ to indicate a particular module is offloaded or not. We can see that in both cases, the offloading policies for sequential DAG only offload one subsequence. Besides, when the extreme events happen with low probabilities, the offloading policy will offload longer subsequence. This is because the better radio conditions, the less task modules will be executed locally.

\begin{figure}[htb]
 \begin{center}
  \begin{tabular}{cc}
    \includegraphics[width= 0.45\columnwidth]{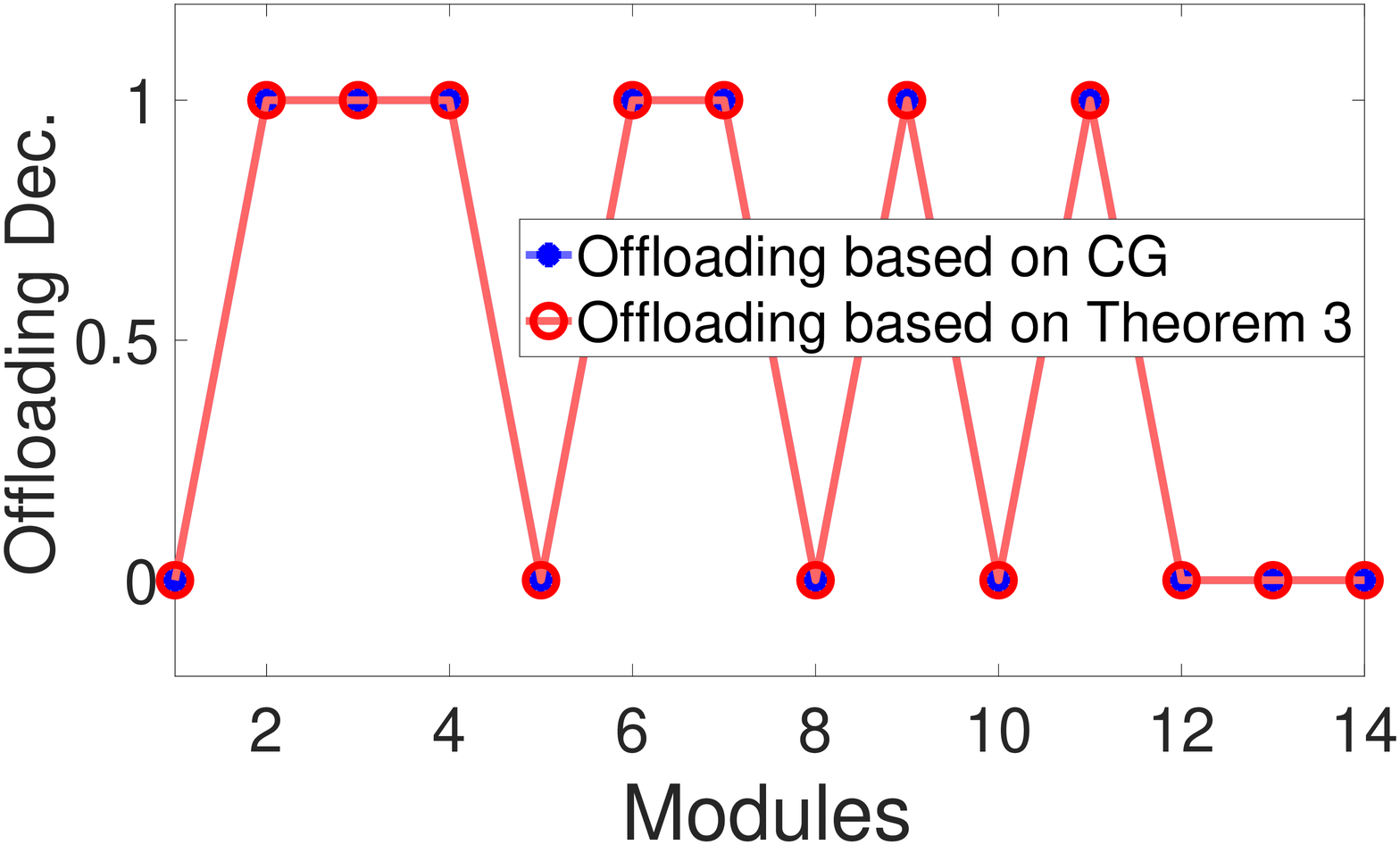}&
         \includegraphics[width= .45\columnwidth]{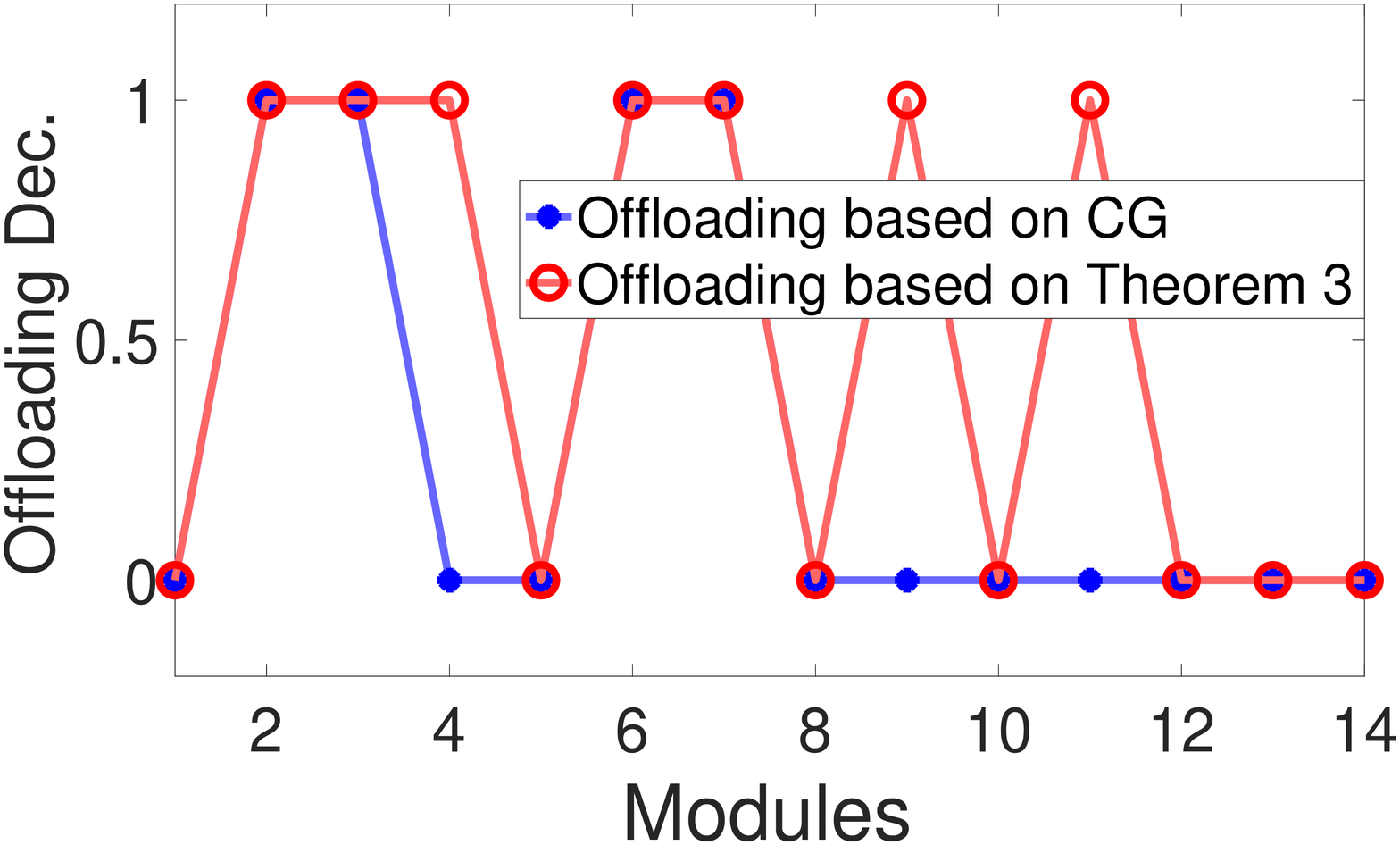}
		\\
		    {\small  (a) $\epsilon=0$, $\epsilon_m^d=\epsilon_m^u=0.01$} 
&
    {\small  (b) $\epsilon=0.03$, $\epsilon_m^d=\epsilon_m^u=0.1$}
        \end{tabular}
      \end{center}
 \caption{\label{fig:para_ex} Offloading decision for parallel dependency DAG.}
\end{figure}


Fig. \ref{fig:para_ex} shows the offloading decision of all tasks with only parallel dependency. We compare the offloading decisions by considering two sets of approximation parameter and probabilities, i.e., $\epsilon=0$, $\epsilon_m^d=\epsilon_m^u=0.01$ and $\epsilon=0.03$, $\epsilon_m^d=\epsilon_m^u=0.1$. In Fig. \ref{fig:para_ex}, the blue dotted lines indicate the offloading policy decided by running our column generation based algorithm, the red circled lines indicate the policy decided by Theorem \ref{para_dep}. We can see that in Fig. \ref{fig:para_ex}(a), the obtained offloading polices are identical for both methods, but in Fig. \ref{fig:para_ex}(b), the policy decided by Theorem \ref{para_dep} offloads more nodes than that decided by the   algorithm. The reason is that when the approximation parameter is large, e.g., $\epsilon=0.03$ in Fig. \ref{fig:para_ex}(b), the solution to the proposed algorithm is an $\epsilon$-bounded approximate solution, while the solution given by Theorem \ref{para_dep} is the optimal offloading decision. This implies that when applications have only parallel dependency, we can directly obtain the optimal offloading policy by evaluating the energy consumption and expected worst-case energy consumption for data transmission of each task module.

Additionally, we also conduct experiments to offload the simulated DAGs with special dependency structures under the uncertainty MEC system described in Section \ref{expsetup}. The   communication parameters  are still set as   those listed in Table \ref{para}. In Table \ref{table:special_compare}, we compare the energy consumption achieved by our scheme, under different choice of $\epsilon$,  $\epsilon_m^d$, and $\epsilon_m^u$, with Hermes and JSCO. Clearly, our developed scheme can still achieve the lowest local energy consumption, which suggests again that by explicitly addressing the uncertainties involved in computation offloading using the extreme value theory, we can make offloading decisions that are close to the optimal ones (i.e., when $\epsilon=0$).  

\begin{table}[h]
\caption{Comparison of energy consumption caused by our algorithm and other schemes on DAG with Special Dependency.} 
\centering 
\begin{tabular}{c|c| c| c| c } 
\hline 
&  \multicolumn{2}{c|}{  sequential DAG}                   &  \multicolumn{2}{c}{   parallel DAG}          \\ [0.5ex] 
    \cline{2-5}
& $\epsilon=0$ & $\epsilon=0.03$ & $\epsilon=0$ & $\epsilon=0.03$\\
\hline
$\epsilon_m^d=\epsilon_m^u=0.01$ & 0.011 J & 0.014 J & 0.017 J  & 0.018 J \\
\hline
$\epsilon_m^d=\epsilon_m^u=0.1$ & 0.016 J & 0.018 J & 0.020 J  & 0.022 J \\
\hline
&  \multicolumn{2}{c|}{0.019 J by Hermes}                   &  \multicolumn{2}{c}{ 0.022 J by Hermes}          \\ 
&  \multicolumn{2}{c|}{0.020 J by JSCO}                   &  \multicolumn{2}{c}{0.024 J by JSCO}          \\ 
    \hline
\end{tabular}
\label{table:special_compare} 
\end{table}

\section{Conclusions}
\label{con}

In this paper, we  investigate the problem of energy-efficient computation offloading in a practical MEC systems. In contrast to existing works which usually impose strong assumptions on communication channels and network queues, we remove these assumptions and address the uncertainties in the MEC system. First, we handle the uncertainties and bound the occurrence probability of extreme events using extreme value theory. Then, we formulate the expected energy consumption in the worst case when executing time-sensitive applications. Next, since the formulated optimization problem is a quadratically constrained binary quadratic programming, we develop an $\epsilon$-bounded approximation algorithm based on column generation technique to solve it. In addition, we also tailor the proposed computation  offloading scheme to accommodate special  applications with only sequential or parallel task module dependency.  We implement the proposed scheme on the Android platform and conduct extensive experiments by executing a real-world mobile application. Experiment results corroborate the fact that it will lead to lower energy consumption by explicitly consider the uncertainties in  communication channels and network queues, and show that our  scheme outperform state-of-the-art schemes in terms of energy saving.

\balance

\bibliographystyle{IEEEtran}
\bibliography{sigproc}

\end{document}